\documentclass[12pt,titlepage]{utarticle}
\usepackage{amssymb,graphicx}
\usepackage{amsmath}
\usepackage{hyperref} 
\usepackage{cite}
\usepackage{upgreek}

\def\e{{\rm e}}

\def\eps{\epsilon}
\def\d{\partial}
\def\l{\left(}
\def\r{\right)}

\newcommand{\be}{\begin{equation}}
\newcommand{\ee}{\end{equation}}
\newcommand{\bea}{\begin{eqnarray}}
\newcommand{\eea}{\end{eqnarray}}
\newcommand{\bg}{\begin{gather}}
\newcommand{\eg}{\end{gather}}
\newcommand{\bseq}{\begin{subequations}}
\newcommand{\eseq}{\end{subequations}}

\renewcommand{\ln}{\mathop{\rm ln}\nolimits}

\newcommand{\Tr}{{\rm Tr}}

\begin{document}
\baselineskip 15.5pt
\title{Solving the Simplest Theory\\[0.3cm] of Quantum Gravity}

\author{Sergei Dubovsky
   \address{
      Center for Cosmology and Particle Physics,\\ Department of Physics,\\
      New York University\\
      New York, NY, 10003, USA\\
   }
   , Raphael Flauger $^{1,}$
   \address{
      School of Natural Sciences,\\
      Institute for Advanced Study,\\
      Princeton, NJ 08540, USA\\
      {~}\\[.5cm]
      {\rm {\it e}-mail}\\[.1cm]
      \emailt{dubovsky@nyu.edu}\\
      \emailt{flauger@nyu.edu}\\
      \emailt{vg629@nyu.edu}\\
   } and Victor Gorbenko $^1$
}

\Abstract{
We solve what is quite likely the simplest model of quantum gravity, the worldsheet theory of an infinitely long, free bosonic string in Minkowski space. 
Contrary to naive expectations, this theory is non-trivial. We illustrate this by constructing its exact factorizable $S$-matrix. Despite its simplicity, the theory exhibits many of the salient features expected from more mature quantum gravity models, including the absence of local off-shell observables,  a minimal length, a maximum achievable (Hagedorn) temperature, as well as (integrable relatives of) black holes. 
All these properties follow from the exact $S$-matrix. We show that the complete finite volume spectrum can be reconstructed analytically from this $S$-matrix with the help of the thermodynamic Bethe Ansatz. We argue that considered as a UV complete relativistic 2-dimensional quantum field theory the model exhibits a new type of renormalization group flow behavior, ``asymptotic fragility". Asymptotically  fragile flows do not originate from a UV fixed point. \vskip 5mm}

\maketitle
\tableofcontents

\section{Introduction and Summary}

Reconciling quantum mechanics with gravity is notoriously hard. The most successful framework to achieve this, string theory, appears very different from quantum field theory and technically considerably more challenging. While chances for direct experimental probes remain slim, a huge amount of theoretical data collected over the years has led to considerable intuition for the properties expected of any theory of quantum gravity. One of these properties, holography as made precise by the AdS/CFT correspondence \cite{Maldacena:1997re,Gubser:1998bc,Witten:1998qj}, rather remarkably suggests that at the end of the day the description of gravity is likely possible through the familiar language of quantum field theory, albeit in an unconventional way.

To understand quantum field theory itself, it proved extremely useful to identify the basic building blocks of the theory, the harmonic oscillator which happens to be the simplest solvable quantum system. This simple building block is then used to construct more sophisticated models.

Needless to say, it would be very helpful to construct such a system for quantum gravity. This appears hard due to the incredible amount of complexity exhibited in gravitational systems. According to string theory, these basic building blocks are strings. While all harmonic oscillators are the same, strings come in many shapes and sizes, some much harder to understand than others. In some regimes other objects, such as membranes \cite{Witten:1995ex} or matrices \cite{Banks:1996vh} are more suitable degrees of freedom to describe the string dynamics. 

In this paper, we study a particularly simple class of these building blocks, the infinitely long critical bosonic string in Minkowski space.\footnote{We will see later that our proposal actually extends beyond conventional critical strings.}
We will show that the worldsheet theory of a single infinitely long string is surprisingly rich even for zero string coupling $g_s=0$. Perhaps even more surprisingly, it behaves like a gravitational theory in a sense that we will make more precise.




In section~\ref{sec:exact}, we show that the theory is non-trivial by extracting its static-gauge $S$-matrix from the known finite volume spectrum. The procedure we use is the same as the one used to extract elastic scattering amplitudes for pions from lattice QCD data \cite{Luscher:1986pf,Luscher:1990ck}. The scattering turns out to be purely elastic, {\it i.e.} the full $S$-matrix is factorizable and reflectionless. In other words the only effect in $2\to2$ scattering is a flavor-independent phase shift which can be extracted from the known two-particle binding energy for the theory on a circle. One finds
 \be
 \label{S-matrix}
 \e^{2i\delta_{NG}(s)}=\e^ {i s\ell_s^2/4}\;,
 \ee
 where $\ell_s$ is the string scale, and $s$ is the conventional Mandelstam variable. For multi-particle scattering processes the phase shift is given by the sum of the two-particle phase shifts.

Given the exact Lorentz-invariant, unitary, analytic and crossing symmetric $S$-matrix, we need a criterion to decide whether it corresponds to a conventional field theory or to a theory of quantum gravity. It is tempting to think that reparametrization invariance is the defining property of gravity. Our starting point, the worldsheet Nambu-Goto theory, satisfies this condition. However, reparametrization invariance, as any gauge symmetry, is just a convenient redundancy in the description. So its presence cannot be used as a sharp criterion (see, e.g., \cite{ArkaniHamed:2007ky}, for a related discussion). In particular, reparametrization invariance does not manifest itself at the level of the $S$-matrix. Nevertheless, one might hope that there is some sense in which reparametrization invariance can be used as a criterion for a theory to be gravitational. In four dimensions  this is made precise by the Weinberg-Witten theorem~\cite{Weinberg:1980kq}. It states that if a massless spin-2 particle is present in the spectrum, the theory cannot have a Lorentz-covariant, ordinarily conserved energy-momentum tensor. The former property is manifest at the $S$-matrix level, and the latter may be considered as a meaningful reformulation of what a reparametrization invariant theory is.

Unfortunately, it is impossible to apply this result directly in a two-dimensional theory. A massless spin-2 field does not have propagating degrees of freedom in two dimensions, so there is nothing to look for at the level of the $S$-matrix. Yet it appears instructive to take the Weinberg--Witten theorem as a guide, and use the absence of the energy-momentum tensor as part of the definition of a gravitational theory in two dimensions. 
Usually, one goes further and assumes that gravitational theories do not have any sharply defined local observables. Heuristically, this is motivated by the difficulty to perform a local measurement with arbitrary precision---the gravitational field of the apparatus eventually causes the gravitational collapse of an observer and of her lab.

We provide evidence for the absence of local observables in the Nambu--Goto theory for the critical string. Generically, the Nambu-Goto theory is non-renormalizable and can only be treated as an effective field theory. This effective field theory breaks down at an energy scale set by the string tension and is UV completed by a field theory, {\it e.g.} a gauge theory or an Abelian Higgs model. In this case, local observables exist and the theory displays no signs of gravitational physics. However, in the critical number of dimensions the Nambu-Goto theory becomes more than an effective field theory. Its exact finite-volume spectrum as well as its $S$-matrix can be calculated and the associated UV behavior is very different from that in a conventional quantum field theory. Direct perturbative calculations as well as non-perturbative arguments sketched in section~\ref{sec:loop} strongly suggest that off-shell correlation functions of local operators nevertheless remain undetermined in this theory.

There is additional evidence for the absence of a well-defined energy-momentum tensor in this theory. Usually, one of its principal characteristics---the UV central charge, which determines the short distance asymptotic of the two-point function, can be extracted from the small radius behavior of the Casimir energy \cite{Cardy:1984rp}. However, as a consequence of the closed string tachyon, the Nambu--Goto Casimir energy becomes complex for radii smaller than the string length, which makes this procedure inapplicable. 

In section~\ref{sec:Bethe} we elaborate on this observation. As a warm-up, we rederive the finite volume vacuum energy from the $S$-matrix using Thermodynamic Bethe Ansatz (TBA) techniques \cite{Zamolodchikov:1989cf}. We then show that the thermodynamic Bethe Ansatz in this case even allows us to analytically extract all excited state energies. This provides a very compelling consistency check for the exact $S$-matrix.
We then exploit the equality of the finite volume Casimir energy and the finite temperature free energy to discuss the thermodynamics of the theory. 
The theory exhibits a maximum temperature, at which both the heat capacity and its integral diverge. Of course, this is nothing but the Hagedorn temperature, well familiar to string theorists, but it is somewhat unusual as a property of a ``conventional" relativistic field theory.
The reader who finds the subsections devoted to the thermodynamic Bethe Ansatz too technical may skip them without loss of continuity and go directly to the discussion of the thermodynamics, however, at the cost of missing an amusing derivation of the spectrum of excitations on a long string.

The exact $S$-matrix~(\ref{S-matrix}) itself is also quite unusual. In section~\ref{sec:packets} we discuss some of its consequences. It does not exhibit any poles, but a cut discontinuity that extends all the way from zero to infinity, featuring an infinite number of broad periodic resonances.

To understand its implications, it is instructive to consider the result of scattering of two Gaussian wave packets. There is no particle production, so the outgoing state is again a pair of Gaussian wave packets. Nevertheless, there are a number of interesting effects. First, the widths of the two outgoing wave packets obey
 \be
 \Delta x_L \Delta x_R\geq \ell_s^2\;.
 \ee
It is natural to  call this the string uncertainty principle. It provides another concrete sense in which the theory does not allow local measurements. Any attempts to resolve space-time details smaller than the string scale are doomed. Interactions make all space-time probes fuzzy at the scale $\ell_s$.

The second surprising effect is a time delay in the center of momentum frame proportional to the energy of the collision,
\be
\label{TimeDelay}
\Delta t=\frac12E\ell_s^2\;.
\ee
This time delay is universal as one would expect in a gravitational theory respecting the equivalence principle. Whether one considers a single hard particle carrying all the energy, or a collection of soft quanta with the same total energy, the time delay remains the same. Such a time delay is suggestive of black hole formation and evaporation. For comparison, in Einstein gravity in four dimensions, black holes result in a universal time delay given by the Hawking evaporation time $\Delta t_H\sim E^3\ell_{Pl}^4$, where $\ell_{Pl}$ is the Planck length. Interestingly, the black hole evaporation time in a more sophisticated example of two-dimensional quantum gravity---the CGHS model \cite{Callan:1992rs,Russo:1992ht,Fiola:1994ir} (see \cite{Ashtekar:2010hx} for a recent update)---is also proportional to 
 the black hole mass, or, equivalently, to the collision energy.
 
Furthermore, outgoing macroscopic packets are highly entangled with each other. The corresponding entanglement entropy for $\ell_s^2\Delta p_L\Delta p_R\gg1$ is 
 \[
 S_{ent}=\log( \ell_s^2\Delta p_L\Delta p_R)\;.
 \]
 
These properties strongly suggest that the amplitude (\ref{S-matrix}) represents an integrable approximation to the process of black hole formation and evaporation. Integrability prevents particle production making it impossible to obtain thermal spectra. However, other expected features, including the high degree of entanglement responsible for the  information ``paradox"  in higher dimensions \cite{Hawking:1976ra}, are present. It is tempting to identify the resonances in our amplitude with the microstates of the black hole.
 
 Finally, in section~\ref{sec:classical} we illustrate that the classical worldsheet theory shares several similarities with more mature gravitational theories as well. To illustrate this, we consider solutions which one might interpret as black hole precursors and use them to explain how the time delay (\ref{TimeDelay}) arises classically. This naturally leads us to discuss the relation between the $S$-matrix determined by~\eqref{S-matrix} and the more familiar (trivial) worldsheet $S$-matrix. To summarize, the time delay~\eqref{TimeDelay} and the $S$-matrix determined by~\eqref{S-matrix} are measured by observers freely falling with respect to the target space, while the trivial worldsheet $S$-matrix is measured by observers freely falling with respect to the worldsheet. The two classes of observers are related by a gauge transformation, and it may seem surprising that they measure different $S$-matrices. However, the $S$-matrix is only invariant under gauge transformations which act trivially at infinity. We will show that the gauge transformation relating the two $S$-matrices does not fall into this class so that there is no contradiction. 
We also present a cosmological solution.
 
 Before proceeding, let us stress that the relation between string theory and two-dimensional quantum gravity is certainly not new (see, e.g., \cite{Ginsparg:1993is} for a review).
 In a sense any two-dimensional theory of quantum gravity can be interpreted as a (non-critical) string theory. However, the emphasis is usually made on  non-critical theories with non-zero string coupling constant. To the best of our knowledge the exact $S$-matrix (\ref{S-matrix}) of a ``free" critical string has not been discussed before, and we feel that the viewpoint advocated here may be useful. We present further speculations and future directions in the concluding section~\ref{conclusions}.

\section{Exact $S$-Matrix of the Critical Nambu--Goto}
\label{sec:exact}
For our purposes it will be instructive to consider the worldsheet theory from an effective field theory point of view. A detailed introduction 
to this approach can be found in the accompanying paper
\cite{effective}. From this point of view the world-sheet theory of an infinitely long string in a $D$-dimensional Minkowski space is a theory of Goldstone bosons corresponding to the coset $ISO(D-1,1)/ISO(1,1)\times SO(D-2)$. Here $ISO(D-1,1)$ is the non-linearly realized target space Poincar\'e symmetry. Its linearly realized subgroup is a direct product of the worldsheet Poincar\'e symmetry $ISO(1,1)$ and of the $SO(D-2)$ group of transverse rotations.
This is a consistent effective field theory in any number of dimensions with a cutoff scale set by the string length $\ell_s$, which physically corresponds to the string width. The effective action starts with the Nambu--Goto term
and in principle has an infinite number of higher derivative corrections, corresponding to higher order geometric invariants.
 
 Somewhat miraculously, the Nambu--Goto theory is expected, at least in a certain sense, to be renormalizable in the critical number of dimensions $D=26$ \cite{Polchinski:1991ax}.
 An effective field theorist would discover this by calculating loops and finding that divergences, which were expected on the basis of the naive power counting, cancel. We will discuss some aspects of these expected cancellations in section~\ref{sec:loop}. We will argue that the story is somewhat subtle. In particular, the cancellations occur only for on-shell quantities. This makes it challenging to see the cancellations by a direct calculation because at low orders in perturbation theory on-shell divergences cancel because of symmetry. To see non-trivial cancellations one thus has to go rather far in the loop expansion.
 
For now we take a shortcut, and do not check the cancellations by brute force calculation. Instead, we deduce the properties of the resulting finite on-shell amplitudes from the known spectrum of the theory at finite volume. This is known for instance from the quantization in light-cone gauge (which is consistent with the non-linearly realized $ISO(D-1,1)$ symmetry at $D=26$). After compactification on a circle (see, e.g., \cite{Polchinski:1998rq}), 
\be
\label{LCspectrum}
E_{LC}(N,\tilde N)=\sqrt{{4\pi^2(N-\tilde{N})^2\over R^2}+{R^2\over \ell_s^4}+{4\pi\over \ell_s^2}\l N+\tilde{N}-{D-2\over 12}\r}\;.
\ee
Here $R$ is the length of the string, $N$ and $\tilde{N}$ are levels of an excited string state, so that $2\pi(N-\tilde{N})/R$ is the total Kaluza--Klein momentum of the state. 

To avoid confusion, let us clarify the meaning of the subscript $LC$. It indicates that we use light-cone quantization to define the theory at the quantum level. However, equation~\eqref{LCspectrum} corresponds to target space energies obtained in light-cone quantization and should not be confused with the spectrum of the light-cone Hamiltonian.
Classically, the target space energy is the same as the energy in static gauge. At the quantum level the definition of the theory in static gauge is ambiguous, and we use~\eqref{LCspectrum} to define the static gauge theory at the quantum level.

One might be skeptical about the existence of an $S$-matrix in the infinite volume limit. Massless two-dimensional theories are often plagued by IR divergences preventing them from providing a well-defined $S$-matrix. Even for massive theories in two-dimensions the kinematics of a scattering process is somewhat subtle. With two and higher spatial dimensions the spatial infinity is connected. A typical scattering process starts from a bunch of particles (wave-packets) incoming from different directions towards the interaction region, and results in a bunch of reaction products escaping at different angles. Both incoming and outgoing states are naturally separated in space and non-interacting at very early and very late stages. 

In the lineland, instead,  the spatial infinity consists of two disconnected points, the ``left infinity" and the ``right infinity".   A scattering process  starts with a collection of right-moving particles at the left infinity and left-moving particles at the right infinity. The outcome is a bunch of left-movers at the left infinity and a bunch of right-movers at the right infinity. Directional separation is not possible any longer. However, if particles are massive, they naturally get segregated according to their propagation velocities, resulting in well-defined free scattering states. 

This argument indicates that one should worry about the existence of an $S$-matrix in the presence of massless particles. Their propagation velocity is energy independent, so that the scattered particles ``sit" on top of each other forever, making it seemingly impossible to define the $S$-matrix. In addition, a single massless particle can split into an arbitrary number of other massless particles propagating in the same direction. 
These effects typically result in soft and collinear divergences plaguing $S$-matrix elements for massless two-dimensional theories.

Nevertheless, a number of examples of well-defined two dimensional massless $S$-matrices are known to exist (see, e.g., \cite{Fendley:1993jh} for a review).
The cleanest examples avoiding these issues arise when the interactions involving only left-movers (or only right-movers) vanish identically. In this case $1\to many$ splittings are absent, and even though left-movers (right-movers) remain on top of each other, they do not interact as soon as there are no right-movers (left-movers) around.
A simple nice example is the theory of a single massless fermion (goldstino) with interactions fixed by a non-linearly realized supersymmetry~\cite{Zamolodchikov:1991vx} (see \cite{Klassen:1989ui} for more examples).

Strings avoid IR problems in this way as well. At tree level it is immediate to check that the on-shell amplitude for $1\to 3$ processes vanishes. Related to this, as discussed in \cite{effective}, the logarithmic terms in the 1-loop $2\to 2$ amplitude, which would lead to IR divergences, all vanish on-shell. 
A non-perturbative argument for the absence of IR divergences in our case can be given as follows~\cite{Zamolodchikov:1991vx}. Consider some number of left-moving particles. They can be made arbitrarily soft by an appropriate Lorentz boost. On the other hand, as a consequence of the shift symmetry, all the interactions are irrelevant in the IR, so that soft particles are free. Consequently, there are no interactions involving left-moving particles only. 

This argument relies only on world-sheet Lorentz invariance and shift symmetry and thus applies to a broad class of effective string theories. For the critical $D=26$ theory similarities with the models discussed in \cite{Zamolodchikov:1991vx},\cite{Klassen:1989ui} go further. Scattering in these models is purely elastic (reflectionless).
The $S$-matrix is diagonal and is completely determined by the phase shift $\e^{2i\delta(s)}$ in $2\to 2$ scattering. 

This is exactly what one expects given the finite volume spectrum (\ref{LCspectrum}). This spectrum implies that the states with a fixed number of particles are exact eigenstates of the Hamiltonian, implying the absence of particle production. Furthermore, different $SO(D-2)$ multiplets with the same number of particles are exactly degenerate, implying the absence of annihilations (and, by crossing symmetry, reflections).
Intuitively, the latter property implies that a string initially oscillating in one direction will keep oscillating in this direction forever. As demonstrated in~\cite{effective} this property holds at tree-level in the Nambu--Goto theory for a relativistic string in any number of dimensions, but is violated away from the critical number of dimensions at one-loop. 

As explained in \cite{Zamolodchikov:1991vx}, the requirements of unitarity, crossing symmetry and  analyticity restrict the phase shift for the purely diagonal massless scattering to take the form
\be
\label{Zamolodchikov}
\e^{2i\delta(s)}=\prod_j{\mu_j+s\over \mu_j-s}\e^{iP(s)}\;,
\ee
where $P(s)$ is an odd polynomial in $s$ and $\mu_j$ are located in the lower half of the complex plane, and either lie on the imaginary axis or come 
in pairs symmetric with respect to it.
The expression  (\ref{Zamolodchikov}) holds for $\mbox{Im}\, s>0$. For $s$ in the lower half of the complex plane the same expression applies with $s$ replaced by $-s$.

The standard expectation is that $P(s)=0$, so that the scattering amplitude is exponentially bounded. Exponential boundedness plus analyticity is commonly taken as the only sharp definition of locality in quantum theories. In agreement with this expectation, the goldstino model of \cite{Zamolodchikov:1991vx} does have $P=0$ and realizes the simplest possible amplitude of this type of the form 
\be
\label{goldstino}
\e^{2i\delta_\text{Gold}(s)}={iM^2-s\over iM^2+s}\;,
\ee
where $M$ is the scale of supersymmetry breaking.

If the critical NG theory indeed has a well-defined $S$-matrix it should also be of the form (\ref{Zamolodchikov}) (times a 
unit matrix in ``flavor" space), but what are the corresponding $\mu_j$ and $P$?

Fortunately, it is straightforward to answer this question. Indeed, the exact spectrum of the theory at finite volume is known and is given by equation~(\ref{LCspectrum}). 
Deducing the scattering amplitudes from the finite volume spectrum is a routine problem in lattice calculations, and the corresponding techniques were developed in 
\cite{Luscher:1986pf}.
Theories in one spatial dimension were specifically  considered  in \cite{Luscher:1990ck}.
%
For the sake of completeness let us sketch a semi-rigorous argument leading to the desired result. 

Consider a two particle eigenstate of the Hamiltonian on a cylinder with a zero total KK momentum, {\it i.e.} with $N=\tilde{N}$ in 
the string case (see (\ref{LCspectrum})). On the one hand, in the Schr\"odinger picture this state $|N,N,t\rangle$ evolves  in time according to
\be
\label{simplephase}
|N,N,t\rangle=\e^{-i E_{LC}(N,N,R)t}|N,N,0\rangle\;.
\ee

On the other hand, we can think of the quanta as traveling around the circle with the speed of light acquiring an additional ${2\delta(s)}$ phase shift due to interactions with a periodicity $\Delta t=R/2$. This implies that in the limit $R,t\to\infty$ the total phase shift can be presented in the form
\be
\label{largeRphase}
|N,N,t\rangle=\e^{-i\l 2 \Delta E_{LC}(N,0,R)+E_{LC}(0,0,R)-4\delta(s)R^{-1}\r t}|N,N,0\rangle\;,
\ee
where $E_{LC}(0,0,R)$ is the ground state energy, and $\Delta E_{LC}(N,0,R)$ is the energy of the one-particle state relative to the vacuum,
\[
\Delta E_{LC}(N,0,R)=E_{LC}(N,0,R)-E_{LC}(0,0,R)\;.
\]
By comparing the two expressions (\ref{simplephase}) and (\ref{largeRphase}) we arrive at the following result for the scattering phase shift,
\be
\label{ourresult}
2\delta(s)=-\lim_{R\to \infty} {R\over 2} \l E_{LC}(N,N,R)-E_{LC}(0,0,R)-2 \Delta E_{LC}(N,0,R)\r\;,
\ee
{\it i.e.}, that the scattering phase is simply proportional to the binding energy of the two-particle state in the large volume limit. 
The square of the center of mass energy $s$ is defined here through
\be
\label{slimit}
s=\lim_{R\to\infty} (E_2(R)-E_0(R))^2\;.
\ee
It follows from (\ref{slimit}), that in order to keep the center of mass energy fixed in the $R\to \infty$ limit, one should also take the limit $N\to \infty$ in such a way that the $N/R$ ratio  is kept fixed, so that
\[
s={16 \pi^2 N^2\over R^2}\;.
\]
From (\ref{ourresult}) we then obtain the expression (\ref{S-matrix}) of the critical NG theory\footnote{ Superficially, the expression (\ref{ourresult}) differs from the one in \cite{Luscher:1990ck} due to the presence of an additional one-particle energy term in (\ref{ourresult}) as compared to that in \cite{Luscher:1986pf,Luscher:1990ck}. However, both produce the same result for $\e^{2i\delta(s)}$, because in the  $R\to \infty$ limit with   $N/R$ ratio being fixed the one-particle energy takes the form
 \[
 \Delta E_{LC}(N,0,R)={2\pi N\over R}\l 1+{\pi (D-2)\ell_s^2\over 6 R^2}+\dots\r\;.
 \]
 }.
 
 The above derivation can be made more rigorous by considering wave-packets and we have checked that this leads to the same answer.  
Note also, that the arguments of \cite{Luscher:1986pf,Luscher:1990ck} were presented for massive particles, when the corrections to the spectrum coming from  the loops wrapping around the circle are 
exponentially suppressed. One may worry that these effects may spoil the derivation in the case at hand, where all particles are massless and these corrections are only power-law suppressed. However,
as a consequence of the shift symmetry, at the case at hand this effect does not affect the scattering phase (\ref{ourresult}). Alternatively, one can understand this from the fact that it is also possible to derive this $S$-matrix directly in an infinite volume version of light-cone gauge.\footnote{We thank Juan Maldacena for pointing this out to us.} 

As a simple consistency check, let us show that this answer agrees with the perturbative calculations presented in \cite{effective}. For definiteness, consider the $t$-channel configuration, {\it i.e.}  with left-moving  1- and 3-particles  with flavors $i$ and $k$ and right-moving
2- and 4-particles with flavors $j$ and $l$. The perturbative one-loop $S$-matrix at $D=26$ then takes the form,
\be
S_{1-loop}={\mathbf 1}+i\delta^2(k_1+k_2-k_3-k_4)\delta_{ik}\delta_{jl}{\ell_s^4 B_{1-loop}\over s} \;,
\ee
where $k_i=(|p_i|,p_i)$ are the two-momenta, the factor of $s^{-1}$ accounts for the correct $(2E)^{-1/2}$ normalization of external legs, and
\[
{\mathbf 1}\equiv\delta_{ik}\delta_{jl}\delta(p_1-p_3)\delta(p_2-p_4)\;.
\] 
Combining the tree-level and one-loop perturbative results of \cite{effective} we find\footnote{We should warn the reader that this  way to calculate the Taylor
coefficients of the exponential function is definitely not the fastest one, even though one of the most amusing.}
\[
B_{1-loop}={s^2\over 2\ell_s^2}+i{s^3\over 16}\;.
\]
This agrees with the exact answer (\ref{S-matrix}) after one notices that in two dimensions (and again for left-moving 1- and 3-particles and right-moving 2- and 4-particles)
\[
\delta^2(k_1+k_2-k_3-k_4)={1\over 2}\delta(p_1-p_3)\delta(p_2-p_4)\;.
\]

The exact result (\ref{S-matrix}) is quite peculiar. It agrees with the general expectation (\ref{Zamolodchikov}) following from analyticity, unitarity and crossing symmetry, and realizes the simplest amplitude with these properties and non-trivial $P(s)$.
Among amplitudes of the form (\ref{Zamolodchikov}) it provides the mildest possible violation of the exponential boundedness. In fact, the amplitude (\ref{S-matrix}) {\it is} exponentially bounded on the physical sheet.

However, the essential singularity of this amplitude at $s=\infty$ is a reminder of the many peculiar properties of string theory.
In the rest of the paper we will discuss some of the consequences of this singularity, and we will argue that the proper interpretation of this $S$-matrix is that it corresponds to a theory of gravity, rather than to an ordinary relativistic quantum field theory.

Before we proceed, the following important comment is in order. All of the discussion in this section and most other results in the paper remain unchanged away from $D=26$.
In other words, we are considering a family of relativistic two-dimensional theories, labeled by a discrete parameter $D-2$ counting the number of flavors in the theory. One may then consider the finite volume spectrum (\ref{LCspectrum}), or the $S$-matrix (\ref{S-matrix}) as the {\it definition} of these models. The special property of the $D=26$ theory is that it non-linearly realizes target-space Poincar\'e symmetry $ISO(25,1)$ and hence provides a consistent quantization of a Lorentz-invariant string.
It would be very interesting to see how this non-linearly realized symmetry arises directly at the level of on-shell amplitudes (see \cite{ArkaniHamed:2008gz} for a related discussion in the context of the  pion chiral Lagrangian and $N=8$ supergravity). However, for most of our discussion this special property of $D=26$ will not be important.

\section{Thermodynamic Bethe Ansatz and Hagedorn Equation of State}
\label{sec:Bethe}
Massless $S$-matrices of the form  (\ref{Zamolodchikov}) usually correspond to integrable RG flows between different conformal field theories. For example \cite{Zamolodchikov:1991vx}, the goldstino $S$-matrix (\ref{goldstino}) corresponds  to the RG flow \cite{Kastor:1988ef} between the tricritical Ising model in the UV and the Ising model in the IR (or equivalently between the $\mathcal{N}=1$ Wess--Zumino model and the theory of a free massless fermion).
It is natural to ask what the corresponding RG flow is in our case. In the IR we have a theory of $(D-2)$ free bosons, but what about the UV?

A useful tool to address this is the thermodynamic Bethe Ansatz~\cite{Zamolodchikov:1989cf}. It allows to reconstruct the vacuum energy $E_0(R)$ of the theory at finite volume, and from it to extract the UV central charge from the short distance asymptotics of $E_0(R)$  \cite{Cardy:1984rp},
\[
E_0(R)\simeq {\pi c\over 6R}\;,\;\;{\mbox{ as }}\;R\to0\;.\footnote{This expression assumes that the CFT in the UV is unitary so that the conformal weight for the ground state is zero. Given that our $S$-matrix is unitary, this seems like a reasonable assumption.}
\]
Of course, in our case there is no need for the TBA procedure to extract the ground state energy. We know the exact spectrum of the theory at finite volume. However, we think it is instructive to rederive the vacuum energy in this way for two reasons. First, it provides a non-trivial consistency check for our $S$-matrix. Second, we will see that the model provides a rare example of a system where the TBA equations can be solved exactly both for the ground state as well as for the excited states. 

\subsection{Thermodynamic Bethe Ansatz for the ground state}
The  TBA procedure is based on the following two key ideas. First,
for any relativistic theory the finite volume energy $E_0(R)$ and the free energy density $f(T)$
at temperature $T$ are related in the following way,
\be
\label{Efrelation}
E_0(R)=Rf(1/R)\;.
\ee
This follows from the path integral representation for the partition function of the Euclidean theory on a torus with circles of length $L$ and $R$ in the limit of large $L$.
On the one hand, this partition function can be thought of as the thermal partition function for the theory compactified on the circle with length $L$ at temperature $1/R$. In this case the large $L$ limit is the thermodynamic limit. On the other hand, it can be thought of as the thermal partition function for the theory compactified on the circle with length $R$ at temperature $1/L$. In this case the large $L$ limit is the low temperature limit. 

The second idea is to use the asymptotic Bethe Ansatz rather than the path integral to calculate the free energy.
That is, one considers the system compactified on a large circle of length $L$ at finite temperature $1/R$. Introducing the particle number densities $\rho^i_{1L}(p)$ and $\rho^i_{1R}(p)$, the quantization conditions for the allowed right-moving momenta \mbox{$p^{(i)}_{k\,R},\; (k=1,2\ldots, \infty)$} are
\be
p^{(i)}_{k\,R}L+\sum_{j=1}^{D-2}\int_0^{\infty} 2\delta(p^{(i)}_{k\,R},p) \rho^j_{1L}(p) dp= 2\pi n^{(i)}_{k\,R},
\label{quantization}
\ee
where $n^{(i)}_{k\,R}$ are positive integers.  The allowed left-moving momenta $p^{(i)}_{k\,L}$ satisfy the same equation with left- and right-movers interchanged. These equations receive corrections at finite $L$, but become exact in the thermodynamic limit. Introducing the level densities  $\rho^i_{L}(p)$ and $\rho^i_{R}(p)$, this becomes the TBA constraint
\be
\label{TBAconstraint}
2 \pi \rho^i_{R}(p)=L+\ell_s^2\sum_{j=1}^{D-2} \int_0^{\infty} p' \rho^j_{1L}(p')dp',
\ee
where we have substituted $2\delta(p^{(i)}_{k\,R},p)=\ell_s^2p^{(i)}_{k\,R}p$, as follows from (\ref{S-matrix}). Notice that this equation implies that the level densities for the long string are in fact independent of flavor and momentum
\be
\rho^i_L(p)=\rho_L\qquad\text{and}\qquad\rho^i_R(p)=\rho_R\,.
\ee
In terms of the level densities and the number densities, the macroscopic energy $H$ and entropy $S$ are 
\begin{gather}
\label{HN}
H=\frac{L}{\ell_s^2}+\sum_{i=1}^{D-2}\int_{0}^{\infty} dp \,p (\rho^i_{1L}(p)+\rho^i_{1R}(p))\,, \\
\hskip-0.9cm S=\sum_{i=1}^{D-2}\int_0^{\infty} dp \left[ (\rho_L+\rho^i_{1L})\log(\rho_L+\rho^i_{1L})-\rho_L \log\rho_L - \rho^i_{1L} \log \rho^i_{1L}\right]\nonumber\\\hphantom{AAAAAAAA}+\sum_{i=1}^{D-2}\int_0^{\infty} dp \left[ (\rho_R+\rho^i_{1R})\log(\rho_R+\rho^i_{1R})-\rho_R \log\rho_R - \rho^i_{1R} \log \rho^i_{1R}\right]\,.
\label{S}
\end{gather}
Notice that the expression for the energy includes the bulk cosmological constant. Regarding the equation for the entropy~\eqref{S}, it is interesting to note that for the long string the particles appearing in the thermodynamic Bethe Ansatz are bosons unlike any other physical examples we are aware of where they are fermions. 

The partition function can then be written as a functional integral over the particle densities
\begin{equation}\label{eq:fi}
Z(R,L)=\int\prod\limits_i\mathcal{D}\rho^i_{1L}\mathcal{D}\rho^i_{1R}\exp\left(-R\, H\left[\rho^i_{1L},\rho^i_{1R}\right]+S[\rho^i_{1L},\rho^i_{1R},\rho_{L},\rho_{R}]\right)\,.
\end{equation}
In the saddle point approximation the integral is dominated by the $\rho_1$ that minimize the free energy \[F[\rho^i_{1L},\rho^i_{1R},\rho_L,\rho_R]=H[\rho^i_{1L},\rho^i_{1R}]-\frac{1}{R}S[\rho^i_{1L},\rho^i_{1R},\rho_L,\rho_R]\] subject to the constraint (\ref{TBAconstraint}).
In terms of the pseudo-energies
\be
\eps^i_L(p)= \frac{1}{R}\log \left( \frac{\rho_L+\rho^i_{1L}(p)}{\rho^i_{1L}(p)} \right)\qquad\text{and}\qquad\eps^i_R(p)= \frac{1}{R}\log \left( \frac{\rho_R+\rho^i_{1R}(p)}{\rho^i_{1R}(p)} \right)\,,
\ee
the equations resulting from the variations of the free energy with respect to $\rho^i_{1L}(p)$ and $\rho^i_{1R}(p)$ are
\begin{align}\label{eq:TBA}
\eps^i_L(p)&=p\left[1+\frac{\ell_s^2}{2\pi R}\sum\limits_{j=1}^{D-2}\int_0^\infty dp'\ln\left(1-e^{-R\eps^j_R(p')}\right)\right]\,,\nonumber\\
\eps^i_R(p)&=p\left[1+\frac{\ell_s^2}{2\pi R}\sum\limits_{j=1}^{D-2}\int_0^\infty dp'\ln\left(1-e^{-R\eps^j_L(p')}\right)\right]\,,
\end{align}
and the free energy at the minimum is
\begin{equation}\label{eq:free}
F=\frac{L}{\ell_s^2}+\frac{L}{2\pi R}\sum\limits_{j=1}^{D-2}\int_0^\infty dp'\ln\left(1-e^{-R\eps^j_L(p')}\right)+\frac{L}{2\pi R}\sum\limits_{j=1}^{D-2}\int_0^\infty dp'\ln\left(1-e^{-R\eps^j_R(p')}\right)\,.
\end{equation}
An immediate consequence of equations~\eqref{eq:TBA} is that the pseudo-energies are independent of flavor and linear in momentum. In addition, by taking an appropriate linear combination of the two equations, one finds that the pseudo-energies for left- and right-movers are identical 
\be\label{eq:epsc}
\eps^i_L(p)=\eps^i_R(p)=p c\,.
\ee
Here $c$ is a solution of the quadratic equation that follows from substitution of equation~\eqref{eq:epsc} into~\eqref{eq:TBA}
\be
c^2-c+\frac{\pi\ell_s^2}{12 R^2}(D-2)=0\,.
\ee
Only the solution that is continuously connected to the one for the free theory, {\it i.e.} the solution that has $c=1$ in the limit of vanishing $\ell_s$ is physical
\be
c=\frac12\left(1+\sqrt{1-\frac{4\pi\ell_s^2}{R^2}\frac{D-2}{12}}\right)\,.
\ee
For this solution, the free energy is
\be
F=\frac{L}{R}\sqrt{\frac{R^2}{\ell_s^4}-\frac{4\pi}{\ell_s^2}\frac{D-2}{12} }\,,
\label{Omega}
\ee
and using the relation (\ref{Efrelation}), one obtains the energy of the ground state 
\be
  E_0(R)=\frac{R}{L} F=\sqrt{\frac{R^2}{\ell_s^4}-\frac{4\pi}{\ell_s^2}\frac{D-2}{12}}\,.
  \label{Egr}
\ee
This agrees perfectly with our starting point, the light cone spectrum (\ref{LCspectrum}) with $N=\tilde{N}=0$. 

\subsection{Thermodynamic Bethe Ansatz for excited states}

TBA equations that allow for an exact solution for the ground state are already rare, but for the long string we can do even better and analytically recover 
the spectrum of all excited states as well in the spirit of~\cite{Dorey:1996re}. To this end, notice that the integrands in the expression for the free energy \eqref{eq:free} and in the TBA equations \eqref{eq:TBA} have branch cuts starting at momenta $p_{k\,L,R}^{(i)}$ such that
\begin{equation}\label{eq:poles}
R\epsilon_{L}^{i}\left(p_{k\,L}^{(i)}\right)=2\pi i n_{k\,L}^{(i)}\qquad\text{and}\qquad R\epsilon_{R}^{i}\left(p_{k\,R}^{(i)}\right)=-2\pi i n_{k\,R}^{(i)}\,.
\end{equation} 
Both $n_{k\,L}^{(i)}$ and $n_{k\,R}^{(i)}$ are a priori arbitrary integers but we inserted a minus sign for right-movers for later convenience.
Deforming the contours in a way that these branch points are circled $N_{k\,L}^{(i)}$ times for left-movers and $-N_{k\,R}^{(i)}$ for right-movers leads to the modified TBA equations
\begin{gather}
\eps^i_L(p)=p+\frac{i}{R}\sum\limits_{j,k}2\delta(p,-i\hat{p}_{k\,R}^{(j)})N_{k\,R}^{(j)}+\frac{1}{2\pi R}\sum\limits_{j=1}^{D-2}\int_0^\infty dp'\frac{d\,2\delta(p,p')}{dp'}\ln\left(1-e^{-R\eps^j_R(p')}\right)\,,\\
\eps^i_R(p)=p-\frac{i}{R}\sum\limits_{j,k}2\delta(p,i\hat{p}_{k\,L}^{(j)})N_{k\,L}^{(j)}+\frac{1}{2\pi R}\sum\limits_{j=1}^{D-2}\int_0^\infty dp'\frac{d\,2\delta(p,p')}{dp'}\ln\left(1-e^{-R\eps^j_L(p')}\right)\,,\label{eq:TBAex}
\end{gather}
where we have introduced $\hat{p}_{k\,L}^{(i)}=-ip_{k\,L}^{(i)}$ and $\hat{p}_{k\,R}^{(i)}=ip_{k\,R}^{(i)}$.\footnote{As we will see, all $\hat{p}_{k\,L,R}^{(i)}$ are real. In particular, this implies that there are no bound states in the theory. The minus sign for left-movers arises because $p_L$ and $\hat{p}_L$ denotes the magnitude of the momentum rather than the momentum itself. There is another natural choice for the double Wick rotation with opposite signs, but it, of course, leads to the same spectrum.} Owing to the relation between the theory compactified on the circle with length $R$ at temperature $1/L$ and the theory compactified on the circle with length $L$ at temperature $1/R$ by double Wick-rotation, we can think of the $\hat{p}_{k\,L,R}^{(i)}$ as momenta of the particles in the theory on the circle of length $R$ that make up the excited state. Their values are determined by~\eqref{eq:poles}
\begin{gather}\label{eq:BAE1}
\hat{p}_{k\,L}^{(i)}R+\sum\limits_{j,m}2\delta(\hat{p}_{k\,L}^{(i)},\hat{p}_{m\,R}^{(j)})N_{m\,R}^{(j)}-i\sum\limits_{j=1}^{D-2}\int_0^\infty \frac{dp'}{2\pi}\frac{d\,2\delta(i\hat{p}_{k\,L}^{(i)},p')}{dp'}\ln\left(1-e^{-R\eps^j_R(p')}\right)=2\pi n_{k\,L}^{(i)}\,,\\\label{eq:BAE2}
\hat{p}_{k\,R}^{(i)}R+\sum\limits_{j,m}2\delta(\hat{p}_{k\,R}^{(i)},\hat{p}_{m\,L}^{(j)})N_{m\,L}^{(j)}+i\sum\limits_{j=1}^{D-2}\int_0^\infty \frac{dp'}{2\pi}\frac{d\,2\delta(-i\hat{p}_{k\,R}^{(i)},p')}{dp'}\ln\left(1-e^{-R\eps^j_L(p')}\right)=2\pi n_{k\,R}^{(i)}\,.
\end{gather}
To underscore the generality of this result, we chose to write these equations in terms of the phase shift $2\delta(p,p')$ but we of course still have $2\delta(p,p')=\ell_s^2p p'$. That equations~\eqref{eq:BAE1} and~\eqref{eq:BAE2} are real for a general phase shift follows from crossing symmetry, which implies that the phase shifts that appear inside the integrals are purely imaginary. In this general form, the massive version of these equations has also appeared {\it e.g.} in the context of the Sinh-Gordon model in~\cite{Teschner:2007ng}. 
In the absence of the third term on the left-hand side, equations~\eqref{eq:BAE1}, \eqref{eq:BAE2} are nothing but the asymptotic Bethe Ansatz equations for the theory on a circle of size $R$ provided both $n_{k\,L,R}^{(i)}$ are positive. The third term together with equations~\eqref{eq:TBAex} provides the finite size corrections. The form of these corrections shows that the third term becomes negligible compared to the second in the thermodynamic limit in which the number of particles goes to infinity as the radius of the circle goes to infinity, showing that equation~\eqref{quantization} indeed becomes exact in this limit as mentioned earlier. 

Finally, the expression for the free energy upon deformation of the contour becomes
\begin{gather}
\hskip-8cmF=\frac{L}{\ell_s^2}+\frac{L}{R}\sum\limits_{j,k}\hat{p}_{k\,L}^{(j)}N_{k\,L}^{(j)}+\frac{L}{R}\sum\limits_{j,k}\hat{p}_{k\,R}^{(j)}N_{k\,R}^{(j)}\nonumber\\\hskip+1cm+\frac{L}{2\pi R}\sum\limits_{j=1}^{D-2}\int_0^\infty dp'\ln\left(1-e^{-R\eps^j_L(p')}\right)+\frac{L}{2\pi R}\sum\limits_{j=1}^{D-2}\int_0^\infty dp'\ln\left(1-e^{-R\eps^j_R(p')}\right)\,.\label{eq:fex}
\end{gather}
To find the solution to this set of equations notice that~\eqref{eq:TBAex} implies that the pseudo-energies are still linear functions of $p$ and independent of flavor. The latter fact is intuitively clear because the interactions are independent of flavor. However, left- and right-movers in general no longer have identical pseudo-energies. The solution is thus of the form 
\begin{equation}
\eps^i_L(p)=c_L p\qquad\text{and}\qquad\eps^i_R(p)=c_R p\,,
\end{equation}  
with $c_L$ and $c_R$ solutions of the system of equations
\bea
c_L=1+\frac{2\pi\ell_s^2\tilde{N}}{c_R R^2}-\frac{\pi\ell_s^2}{c_R R^2}\frac{D-2}{12}\,,\\
c_R=1+\frac{2\pi\ell_s^2N}{c_L R^2}-\frac{\pi\ell_s^2}{c_L R^2}\frac{D-2}{12}\,,
\eea
with \[N=\sum\limits_{j,k}n_{k\,L}^{(j)}N_{k\,L}^{(j)}\text{ and }\tilde{N}=\sum\limits_{j,k}n_{k\,R}^{(j)}N_{k\,R}^{(j)}\,,\] and the momenta are quantized according to \[\hat{p}_{k\,L,R}^{(i)}=\frac{2\pi n_{k\,L,R}^{(i)}}{c_{L,R}R}\,.\]
As before, only the solution that is continuously related to the one for the free theory is physical. For this solution, equation~\eqref{eq:fex} leads to
\begin{equation}
E(N,\tilde{N})=\frac{R}{L}F=\sqrt{{4\pi^2(N-\tilde{N})^2\over R^2}+{R^2\over \ell_s^4}+{4\pi\over \ell_s^2}\l N+\tilde{N}-{D-2\over 12}\r}\,,
\end{equation}
precisely reproducing the light-cone spectrum~\eqref{LCspectrum}. While this is a rather contrived derivation of the spectrum of a string, it provides a compelling consistency check of the $S$-matrix~(\ref{S-matrix}). Even though the quantization condition (\ref{quantization}) is essentially the same equation that we used to derive the
scattering phase shift in section~\ref{sec:exact}, it is applied in a very different regime in the TBA procedure. In section~\ref{sec:exact} we applied it for two-particle states, while the TBA derivation operates in the thermodynamic limit.
To appreciate the difference, note, for example,  that any modification of the spectrum by terms decaying faster than $1/R$ at large $R$ would result in the same $S$-matrix. Such a finite volume spectrum does not pass the TBA cross-check, indicating that it is incompatible with the Lorentz symmetry.

Before moving on let us make one brief remark. We were not too careful in our definition and evaluation of~\eqref{eq:fi} and simply stated that the saddle point approximation amounted to the minimization of the free energy. It seems plausible that in a more careful treatment excited states appear directly as subleading saddle points. This is beyond the scope of this paper, but might lead to a more satisfactory derivation of the thermodynamic Bethe Ansatz for excited states.
  
%
\subsection{Hagedorn equation of state}
For any relativistic theory the finite volume ground state energy determines the equation of state, the free energy as a function of temperature according to equation~\eqref{Efrelation}.
The unusual property of the equation of state for the string is that the free energy becomes complex above certain critical temperature
$$T_H={1\over \ell_s} \sqrt{\frac{3}{\pi(D-2)}}\,.$$
To understand the physical meaning of this let us calculate some basic thermodynamic properties of the system. To reproduce the standard field theory calculation, for this purpose we subtract the cosmological constant from the free energy so that the new free energy vanishes in the limit of zero temperature
\be\label{thermoF}
F(T)=\frac{L}{\ell_s^2}\sqrt{1-\frac{T^2}{T_H^2}}-\frac{L}{\ell_s^2}\,.
\ee

Let us now calculate the heat capacity $c_v$ in the vicinity of the critical temperature $T_H$.
Using the relation between the energy density $\uprho$, pressure $p$ and entropy density $s$ 
$$p=-\uprho+s T\,,$$
the first law of thermodynamics
$$
dp=s dT
$$
and the relation of the pressure to the free energy $p=-F/L$, we find
 \be
 \label{cv}
 c_v = T \frac{\d^2 p}{\d T^2} =\frac{TT_H}{\ell_s^2(T_H^2-T^2)^{3/2}}\sim (T_H-T)^{-3/2}\;.
 \ee
 We see that both the heat capacity and its integral $\int c_v dT$ diverge at the critical temperature. This indicates
that $T_H$ is really the maximum physical temperature, it is impossible to reach it by supplying a finite amount of energy to the system.
Of course, all this is just the familiar Hagedorn behavior of string theory, and indeed this critical temperature $T_H$
is equal to the Hagedorn temperature. However, seen as a property of a ``conventional" relativistic two-dimensional field theory it appears quite unusual.

The knowledge of the exact equation of state allows to calculate further physical properties of the system. 
In particular, with the free energy~(\ref{thermoF}), the energy density can be written as a function of the pressure
\be\label{rofp}
\uprho={p\over 1-\ell_s^2 p}\;.
\ee
The Hagedorn behavior sets in near $p=1/\ell_s^{2}$, where the energy density diverges in agreement with (\ref{cv}).
One finds that the sound velocity,
\be\label{cs}
c_s=\l\frac{\d \uprho}{\d p}\r^{-1/2}=1-\ell_s^2 p\;,
\ee
vanishes when the temperature approaches $T_H$. 

We have subtracted the cosmological constant for the purpose of our thermodynamic calculations to calculate the energy density and pressure a putative static-gauge observer on the string would measure on the walls of his box filled with gas. It is also interesting to consider the result of the calculation including the cosmological constant. This changes the energy density and the pressure by a constant shift with opposite sign \[\uprho_T=\uprho+\frac{1}{\ell_s^2}\qquad\text{and}\qquad p_T=p-\frac{1}{\ell_s^2}\,.\]
While the heat capacity remains unchanged, the speed of sound as a function of the pressure becomes 
\[c_s=-\ell_s^2 p_T\,.\]
and the relation between the energy density and pressure is of the form of the Chaplygin gas
\be\label{Chap}
\uprho_T=-\frac{1}{\ell_s^4p_T}\,,
\ee
At temperatures far below the Hagedorn temperature, $p_T\approx-1/\ell_s^2$ so that $\uprho_T\approx -p_T\approx1/\ell_s^2$. However, as the temperature approaches $T_H$, the pressure from the gas nearly cancels the pressure from the cosmological constant so that $p_T\approx0$ and the unusual behavior of the Chaplygin gas equation of state becomes important.

Let us remark in passing that equations~\eqref{thermoF}-\eqref{Chap} are valid even if one introduces a chemical potential $\mu$ for the number of particles. which is sensible because the theory is integrable. The critical temperature $T_H$ is then determined from
$$T_H={1\over \ell_s} \sqrt{\frac{\pi}{2(D-2)\text{Li}_2(e^{\mu/T_H})}}\,.$$

Returning to our original goal of extracting the value of the central charge from the UV behavior of the Casimir energy, we  have failed. The vacuum energy becomes complex at small $R$, and even if one neglects this and formally expands it around $R=0$ there is no $1/R$ term in this expansion. It is also instructive to consider the behavior of the energy for excited states
\be
E(N,\tilde{N})\sim \frac{2\pi}{R}|N-\tilde{N}|\,.
\ee
This should be contrasted to the behavior in a conformal field theory
\be
E(N,\tilde{N})\sim \frac{2\pi}{R}\left(h+\tilde{h}+N+\tilde{N}-\frac{c}{24}-\frac{\tilde{c}}{24}\right)\,.
\ee
While right- and left-movers are decoupled in the conformal field theory, this is not the case for the theory on the string worldsheet. This can also be seen from the fact that our $S$-matrix between right- and left-movers~\eqref{S-matrix} does not approach a constant as $s\to\infty$. 
We take this behavior as a strong indication that the exact $S$-matrix (\ref{S-matrix}) does not correspond to a conventional RG flow between a UV and IR fixed point.

\section{Absence of Local Off-Shell Observables}
\label{sec:loop}
The failure to extract the value of the central charge in the conventional way described in the previous section suggests that the energy-momentum tensor does not exist as a local operator in the model at hand. As discussed in the {\it Introduction} this is an expectation (or rather part of the definition) of a quantum theory of gravity. In fact, one expects problems defining any local operator. 
This appears natural if one starts with the reparametrization invariant Nambu--Goto action
\be
\label{NGdiff}
S_{NG}=-\frac{1}{\ell_s^{2}}\int d^2\sigma\sqrt{-\det{\d_\alpha X^\mu\d_\beta X_\mu}}\,.
\ee
Just like in four-dimensional gravity, one can then introduce external sources in a covariant way and define the generating functional for the corresponding local operators.
The resulting Green's functions of  local fields will transform {\it covariantly} rather than remain {\it invariant} under reparametrizations. Hence they cannot correspond to physical observables, which should be gauge invariant.

That this argument is too naive can be seen from the fact that we can start with a  conventional field theory with a well-defined set of local operators and 
introduce reparametrization invariance by promoting the parameters of gauge transformations to dynamical fields. The resulting theory is now reparametrization invariant, but possesses the same set of local operators as the original theory.
 
Similarly, one can fix a gauge in a covariant theory and seemingly get around this objection. One example, discussed in \cite{effective}, is static gauge with $X^0=\tau$, $X^1=\sigma$. From an effective field theory viewpoint this seems a very natural starting point to describe the dynamics of long strings.
In this gauge, Green's functions of local operators can be calculated perturbatively. This is perfectly consistent with the possibility to use the Nambu--Goto actions as an effective description of cosmic strings or flux tubes in conventional field theories, where no problems should arise when defining local observables. 

Similarly, in four dimensional gravity one can perform perturbative calculations for local correlators, and they may be relevant physically, if interpreted properly. 
All this strongly suggests that real problems with defining local observables in gravitational theories start only at the non-perturbative level  (see, e.g., \cite{ArkaniHamed:2007ky} for further discussion).

For the case at hand, the problems with defining local observables are partially related to the fact that the worldsheet theory is non-renormalizable by  naive power counting. Before we turn to the problems with off-shell observables, it is instructive to discuss what happens on-shell.
In the critical number of dimensions, the theory gives rise to a well-defined finite volume spectrum (\ref{LCspectrum}) and  $S$-matrix (\ref{S-matrix}). Presumably, this implies an infinite number of cancellations in the static gauge perturbative calculations. 

Unfortunately, to see non-trivial cancellations is technically somewhat challenging. 
The one-loop on-shell amplitudes calculated in \cite{effective} are finite. However, there is no counterterm which could contribute to on-shell scattering at this order, so this is not a non-trivial check. To see non-trivial cancellation one should calculate at least the two-loop $2\to 2$ scattering. 

Note  that the $S$-matrix (\ref{S-matrix}) does not imply that there are no infinities in the Feynman amplitudes following from the Nambu--Goto action itself.
The $S$-matrix is analytic in $s$. This implies that there can be no infinite tree-level counterterms because those would signal the presence of logarithmic terms $\log (s\ell_s^2)$ in the derivative expansion of the amplitudes. However, at the moment we cannot exclude the presence of {\it finite} counterterms, which contribute to cancellation of infinities through loops. 


The finite counterterms are expected to depend on the renormalization scheme. To understand their structure it may be useful to explicitly construct the infinite set of conservation laws responsible for the factorizability of the $S$-matrix (\ref{S-matrix}). At the classical level these are the symmetries of the Nambu--Goto action. However, any given regularization scheme may break them, introducing finite counterterms needed to compensate for this breaking. As discussed in~\cite{effective}, a somewhat similar situation occurs for the target space $ISO(D-1,1)$ symmetry if one uses Weyl symmetric ordering with $\zeta$-function regularization as renormalization scheme.

We see that from an effective field theory perspective renormalizability of on-shell quantities in the integrable theory (\ref{S-matrix}) is somewhat subtle. However, the absence of off-shell observables in some sense is less subtle. To see why, it is instructive to study a simple example, such as the one-loop matrix element of the $\d X$ operator between the vacuum and a three-particle state with two left-movers $p_1, \, p_2$ and one right-mover $p_3$, with the flavor indices $j, \,k$ and $l$ correspondingly. This amounts to taking one of the legs in the one-loop $2\to 2$ scattering amplitudes studied in \cite{effective} off-shell. An explicit calculation gives
\be
\label{divergence}
\langle  0|\d_\alpha X^i |p_1p_2,p_3\rangle\sim \delta^{il}\delta^{jk} {1\over \pi \epsilon}q_\alpha p_1p_2p_3^2+...\;,
\ee
where we dropped the finite part and $q_\alpha=(p_1+p_2+p_3,p_1+p_2-p_3)$.

 From an effective field theory point of view this infinity is not unexpected, and does not present a challenge. One needs to renormalize the operator $\d_\mu X^i $ by subtracting the corresponding counterterm, for example,
 \[
\left[ \d_\alpha X^i\right]_{ren}=\d_\alpha X^i-{\ell_s^2\over 8 \pi \epsilon} \d_\alpha \l \d_\beta \d_\gamma X^i \d^\beta X^j\d^\gamma X^j \r \;.
\]
 This operator mixing is allowed by the symmetries and does not contribute to on-shell scattering, so there is no contradiction with the previous arguments.

The presence of this mixing indicates that as far as the off-shell quantities are concerned the critical Nambu--Goto theory is just like any other non-renormalizable theory. One can calculate Green's functions of local operators order-by-order in the derivative expansion at the expense of introducing a finite number of new parameters at any given order. The expansion breaks down above the cutoff scale $\ell_s^{-1}$.  This is just another way of saying that the theory  does not predict off-shell observables.
Note that even in renormalizable (and free) theories one encounters additional infinities when trying to define composite operator. However, there any given operator
mixes only with a finite number of other operators, as follows from dimensional analysis. This is different for non-renormalizable theories, where the coupling has a negative mass dimension. In particular, here we find mixing even for the elementary fields $\d X$.

While this is suggestive, it does not prove that off-shell observables do not exist. Most of our arguments apply equally to the theory of the goldstino describing the flow between the tricritical and the critical Ising model. There, however, even though the theory appears non-renormalizable, cancellations are expected to persist even for off-shell quantities.
This can be seen from a complementary point of view which does not rely on the perturbative expansion, but uses powerful techniques \cite{Smirnov:1992vz} which often allow to reconstruct local correlators from the exact $S$-matrix in two-dimensional integrable models.  The idea is to express the local correlators in terms of sums of products of form factors and to make use of the expected analytic dependence of these form factors on momenta.

For the massless case, in particular the theory of the goldstino, the recipe is summarized in \cite{Delfino:1994ea}. It reduces to the following. A form-factor of a local operator $\cal{ O}$ inserted at the origin $\tau=\sigma=0$, takes the following general form,
\begin{gather}
\label{generalform}
\langle 0|{\cal{O}}(0)|p_{L1},\dots, p_{Ll};p_{R1},\dots, p_{Rr}\rangle=Q_{r,l}(\{p_{L}\};\{ p_R\})\times\nonumber \\
\prod_{1\leq i<j\leq l}{1\over p_{Li}-p_{Lj}}\prod_{1\leq i<j\leq r}{1\over p_{Ri}-p_{Rj}} \prod_{1\leq i\leq l\;1\leq j\leq r} f(\log (4\ell_s^{2}p_{Li}p_{Rj}))\;,
\end{gather}
where $p_{Li},p_{Rj}$ are the sets of positive left- and right-mover's momenta. The coefficient functions $Q_{r,l}$ are separately symmetric in $p_{Li}$ and $p_{Rj}$, and  analytic everywhere, possibly apart from $p_{Li},p_{Ri}=0,\infty$.
The function $f(\beta)$ is a minimal solution of the Riemann--Hilbert problem determined by the phase shift (\ref{S-matrix}), {\it i.e.} it has no poles or zeros in the strip $0<{\mbox Im}\beta<2\pi$ and satisfies
\be
\label{Riemann}
f(\beta)=e^{2i\delta_{NG}(4\ell_s^{-2}e^\beta)}f(\beta+2\pi i)\;.
\ee 
The coefficient functions $Q_{r,l}$ satisfy a set of recursion relations, whose explicit form is not needed for our present discussion. One usually expects the function $f$ to scale as a finite power of $e^\beta$ at large $\beta$ as is the case in the goldstino theory (\ref{Zamolodchikov}) considered in \cite{Delfino:1994ea}. Assuming that the form-factors are exponentially bounded, then restricts the $Q_{r,l}$'s to be rational functions. 
Given a specific choice of an operator ${\cal O}$, one can obtain further information about the infrared  $p\to 0$ asymptotics of $Q_{r,l}$'s from perturbative calculations, allowing to completely fix the $Q_{r,l}$'s for a few values $r,l$. The rest is then found by solving the recursion relations.

In our case the situation is different. The minimal solution of (\ref{Riemann}) for the phase shift (\ref{S-matrix}) takes the form
\be
\label{Riemann_sol}
f(\beta)=\exp{\l-\beta e^\beta/2\pi\r}
\ee
At large positive $\beta$ (which correspond to large momenta) this function decays faster than the exponent as a function of the momenta. However, it grows faster than exponentially for $\mbox{Im} \beta=\pi$.
Given this analytic behavior there is no reason to assume that the $Q_{r,l}$'s have polynomial growth. So at least in its standard form, this technique cannot determine the form-factors in our case. 

This still does not constitute a rigorous proof that off-shell local observables cannot be defined in this $1+1$ dimensional field theory, but is merely a manifestation of our inability to extract them.
However, further evidence supporting the claim that they do not exist comes from thinking about all this from the perspective of the full interacting string theory. If one succeeded in defining local off-shell observables in this theory, it would imply that it is possible to introduce local external probes in string theory. This is strongly believed to  be impossible \cite{Polchinski:1998rq}.
The same argument implies that in addition to the $S$-matrix (\ref{S-matrix}) there is a set of well-defined (non-local) observables which are off-shell from the point of view of our $1+1$ dimensional theory. These are conventional string scattering amplitudes, constructed through insertions of vertex operators. We leave the task of studying these in the static gauge language for the future, but discuss the implications of the existence of these observables from the world-sheet perspective in more detail in concluding section \ref{conclusions}.

Before concluding this section, let us address one further subtlety regarding the definition of local operators. One might argue that in light-cone gauge our theory is that of $D-2$ free bosons for which local observables certainly do exist. We will briefly explain why these are in fact not local observables of our theory. 

In any Lorentz-invariant theory, the definition of local observables includes that the corresponding operators in the Heisenberg picture transform according to\footnote{For simplicity the formula is written for scalar operators, but our arguments will not depend on this.}
\be\label{eq:lt}
U(a,\Lambda)\mathcal{O}(x)U^{-1}(a,\Lambda)=\mathcal{O}(\Lambda x+a)\,.
\ee 
Here, as usual $U(a,\Lambda)$ is the unitary operator representing a boost $\Lambda$ generated by $M$ and a translation in space and time generated by $P$ and $H$, respectively. For the theory to be Lorentz invariant, we require that $U(a,\Lambda)$ act in the same way on the Heisenberg picture $in$- and $out$-states
\be\label{eq:inout}
U(a,\Lambda)|p_1,p_2,\dots,\pm\infty\rangle=e^{-i(a_\alpha p_1^\alpha+a_\alpha p_2^\alpha+\dots)}\sqrt{\frac{(\Lambda{p_1})^0}{p_1^0}\frac{(\Lambda{p_2})^0}{p_2^0}\cdots}|\Lambda p_1,\Lambda p_2,\dots,\pm\infty\rangle\,.
\ee
While innocuous at first sight, property~\eqref{eq:inout} contains information about the $S$-matrix and encodes its Lorentz invariance. Generically, there may be more than one representation of the Poincar\'e group acting on the physical Hilbert space, but only the one satisfying~\eqref{eq:inout} is physically meaningful. We thus refer to an observable as local if it is defined at a point and satisfies the transformation property~\eqref{eq:lt} for those $U(a,\Lambda)$ that are compatible with our $S$-matrix~\eqref{S-matrix} in the sense of~\eqref{eq:inout}. 

It can be shown that the light-cone gauge operators $X^i(\sigma)$ fail to satisfy the transformation property~\eqref{eq:lt} for the Poincar\'e group compatible with our $S$-matrix. As a consequence, they are not local operators of our theory.  

The light-cone theory, of course, has its own (trivial) $S$-matrix and the $X^i(\sigma)$ do satisfy~\eqref{eq:lt} with respect to the Poincar\'e group compatible with this $S$-matrix. However, this theory should be considered as a different theory. The light-cone theory and our theory $are$ related by a gauge transformation. So it may seem surprising that they give rise to different S-matrices since the $S$-matrix is usually a gauge invariant observable. However, this only applies for gauge transformations act trivially at infinity. The gauge transformation relating static gauge and light-cone gauge can be shown not to fall into this class and thus $does$ affect the $S$-matrix. We will elaborate on these issues and give a physical interpretation of these coordinates in section~\ref{sec:classical}. 

\section{Quantum Black Holes and String Uncertainty Principle}
\label{sec:packets}
To develop further intuition for the physical properties of the $S$-matrix (\ref{S-matrix}), it is instructive to study its observational consequences in scattering processes.
Given that the $S$-matrix is purely diagonal, there are no observables of the type one is used to in collider experiments in particle physics. There is no particle production, and in two dimensions there is no deflection angle to measure. 
The only non-trivial part of the $S$-matrix is an overall phase.

 To measure the phase one needs to set up an interferometry experiment. For instance,
one can imagine splitting a right-moving wave packet into two and sending them along two different ``string theory" arms. One arm is empty, and another contains a left-moving packet. At the end one brings the two out-going 
right-movers back together and observers the relative time delay, interference pattern, {\it etc.}

Any discussion of this kind necessarily brings in some tension with the previous arguments regarding the absence of off-shell observables. Here we are imagining that one can design such an interferometer, which would imply the possibility of attaching external knobs to the system, which is very much the same as performing an off-shell measurement. This is similar to  separation between classical and quantum in the Copenhagen interpretation of quantum mechanics.
As in the latter case, this procedure is instructive for practical purposes if interpreted with care, even if it carries some intrinsic logical inconsistency.

With these qualifications in mind, let us proceed to the study of two-body scattering characterized by the phase shift (\ref{S-matrix}).
For simplicity, we consider scattering for a two particle state $|L,R\rangle$ with one left-mover and one right-mover of different ``flavors" $X^i$ and $X^j$ ($i\neq j$). In the interaction picture, the $in$-state is 
\be
\label{in-state}
|L,R,t\to-\infty\rangle=\int_0^\infty dp_L\int_0^\infty dp_R \,f_L(p_L)f_R(p_R)\alpha^{i\dagger}(p_L)\tilde\alpha^{j\dagger}(p_R)|0\rangle\;,
\ee 
where $\alpha^{i\dagger}$ and $\tilde\alpha^{j\dagger}$ are creation operators for left- and right-movers, respectively, and $f_L, f_R$ are the corresponding momentum space wave-functions.\footnote{We use conventions such that $[\alpha^i(p),\alpha^{j\dagger}(p')]=\delta^{ij}\delta(p-p')$ and similarly for right-movers. In these conventions the wave functions are normalized as $\int_0^\infty dp|f(p)|^2=1$. }
In writing (\ref{in-state}) we assumed that the particles are initially uncorrelated. This is physically the most interesting case, and it is straightforward to extend the discussion to general states, if needed.
By definition of what we mean by the $S$-matrix, at late times the same state is described by
\be
\label{out-state}
|L,R,t\to\infty\rangle=\int_0^\infty dp_L\int_0^\infty dp_R \,f_L(p_L)f_R(p_R)e^{ip_Lp_R\ell_s^2}\alpha^{i\dagger}(p_L)\tilde{\alpha}^{j\dagger}(p_R)|0\rangle\;.
\ee 
%
%
We can extract the time delay from the probability density for finding the left-moving particle at position $x$ at some time $t$ long after the collision, or equivalently from the diagonal part of its real-space density matrix
\be
\label{coordrho}
\rho(t,x_L,x_L)=\frac{1}{2\pi}\int_0^\infty dp_L\int_0^\infty dp'_L\,\rho(p_L,p'_L)e^{-i(p_L-p'_L)(t+x_L)}\,.
\ee
where $\rho(p_L,p'_L)$ is the density matrix for the left-moving particle in the momentum representation
\be
\label{momentumrho}
\rho(p_L,p'_L)=\int_0^\infty dp_R\,f_L(p_L)f_L^*(p'_L)|f_R(p_R)|^2e^{ip_R(p_L-p_L')\ell_s^2}\;.
\ee
To make the discussion as explicit as possible, let us take the initial wave packets to be Gaussian with average momenta $\bar{p}_{R,L}$ and momentum spreads equal to $\Delta p_{L,R}$. For narrow packets $\Delta p_{L,R}\ll \bar{p}_{L,R}$ the range in the momentum integrals can be extended to the entire real line without changing the result much allowing us to write the result of the integration in simple closed form
\be
\label{scrambled}
\rho(p_L,p'_L)=f_L(p_L)f_L^*(p_L')e^{i\bar{p}_R(p_L-p'_L)\ell_s^2-{1\over 2}(p_L-p'_L)^2\Delta p_R^2\ell_s^4}\;.
\ee
Since this result is rather remarkable in its own right, let us briefly postpone our goal of extracting the time delay to discuss it. For macroscopic  $p_L,p'_L,\Delta p_R$ off-diagonal elements of this density matrix are highly suppressed, indicating that the scattering process results in a highly entangled  state, reminiscent of black hole creation and evaporation in a gravitational theory. As a consequence of integrability, we cannot hope for thermalization in the sense of the canonical ensemble. However, the diagonal form of the density matrix in the energy eigenstate basis is indicative of thermalization in the microcanonical sense. The corresponding high degree of entanglement is exactly what gives rise to the information loss ``paradox" in  higher-dimensional theories \cite{Hawking:1976ra}. 
It is interesting to calculate the entropy associated with the density matrix~\eqref{scrambled}. For the interested reader, we include the calculation of the entanglement entropy in the {\it Appendix}. Here we only give the final result in the limit of large $\Delta p_R\Delta p_L \ell_s^2$
\be
\label{entent}
S_{ent}=\log{\Delta p_R\Delta p_L \ell_s^2}\;.
\ee
The logarithmic dependence of the entropy on the quantum gravity scale $\ell_s$ appears natural, given that the geometric entanglement entropy is logarithmically divergent in two dimensions.

It is straightforward to see that the result (\ref{entent}) does not only apply to $2\to 2$ scattering. Given an arbitrary uncorrelated initial state 
$| in\rangle=|\psi_L\rangle\otimes|\psi_R\rangle$ the worldsheet scattering results in the final state of the form
\be
\label{psiout}
| out \rangle=e^{iP_L P_R \ell_s^2}|\psi_L\rangle\otimes|\psi_R\rangle\;,
\ee
where $P_{L(R)} $ are the operators of the total left(right)-mover's momentum. The scattering phase in (\ref{psiout}) depends only on the total momenta of the initial states. As a result, one can repeat verbatim the above calculation of the entanglement entropy, by separating the center-of-mass coordinates from the internal degrees of freedom in the scattering states
$|\psi_{L(R)}\rangle$. This results in the same expression (\ref{entent}), where $\Delta p_{L(R)}$ stand now for the dispersions in the total momenta. 

The left-left density matrix corresponding to the scattered state (\ref{psiout}) can be written in the following suggestive form
\be
\hat{\rho}(p_L,p'_L)=2\pi |\psi_L(p_L) \rangle\langle \psi_L(p'_L)| \int_{-\infty}^{\infty}\,dx_R\,\psi_R^*(x_R) \psi_R(x_R+\ell_s^2(p_L-p'_L))\;,
\ee
where we have used the notation
\begin{gather}
|\psi_L(p_L) \rangle=\sum\limits_{N_L}\langle N_L,p_L|\psi_L\rangle |N_L,p_L\rangle\,,\\
\psi_R^*(x_R) \psi_R(x_R')=\sum\limits_{N_R}\int_{-\infty}^\infty \frac{dp_R}{2\pi} \int_{-\infty}^\infty \frac{dp_R'}{2\pi} \langle N_R,p_R'|\psi_R\rangle\langle\psi_R |N_R,p_R\rangle e^{-ip_Rx_R}e^{ip_R'x_R'}\,,
\end{gather}
where $|N_{L,R},p_{L,R}\rangle$ are a set of orthonormal states with non-zero eigenvalues of left- and right-moving momenta $p_{L,R}$ that span the left- and right-moving sectors and $N_{L,R}$ includes discrete and continuous labels. As before, we extended the momentum integrals over the entire real axis assuming that the spreads in momentum of the $\langle N_R,p_R|\psi_R\rangle$ are much smaller  than the means.   

We see that off-diagonal elements of the density matrix are suppressed by the matrix element characterizing an overlap of the right-moving state with itself shifted by $\ell_s^2(p_L-p'_L)$, which is tiny for macroscopic states.

Let us now return to the task of extracting the time delay in the $2\to 2$ process from the diagonal part of the density matrix in coordinate representation. For the Gaussian wave packets, we find 
\be
\label{BHrho}
\rho(t,x_L,x_L)=\frac{1}{\sqrt{2\pi}\Delta x_L}\exp\l-{(t+x_L-\bar{p}_R\ell_s^2)^2\over 2\Delta x_L^2}\r\text{ with  }\,\Delta x^2_L={1\over 4}\l\Delta p_L^{-2}+4\ell_s^4\Delta p_R^2\r\;.
\ee
This probability density exhibits two interesting properties. First, the spatial spread of the scattered packet has increased. It is natural to call this the stringy uncertainty principle. It is consistent with the evidence that no local observables exist and prevents one from measuring space-time events with a resolution better than the string length $\ell_s$. This can be made more precise by inspection of the two-particle probability distribution after the collision. One finds that this stringy uncertainty principle can be written in a more suggestive Lorentz-invariant form
\be\label{eq:su}
\Delta x_L\Delta x_R\geq \ell_s^2\,.
\ee
The second distinctive feature of the probability distribution is the large time delay for macroscopic objects experienced by the outgoing wave packet 
\be
\label{evaporation_time}
 \Delta t=\bar{p}_R\ell_s^2\,.
\ee
This is in agreement with the black hole interpretation of the amplitude (\ref{S-matrix}). The Hawking temperature of two-dimensional black holes is independent of the mass, resulting in an evaporation time linear in the mass. It is worth stressing that time delays that grow indefinitely with energy are highly unusual in conventional quantum field theories. 

With the black-hole interpretation, it may be surprising that the evaporation time (\ref{evaporation_time}) for the left-mover depends only on the energy of the right-mover and not on the total center of mass energy of the collision. However, this is perfectly consistent with the black hole interpretation and Lorentz invariance. To see this, note that the two-momentum of a created black hole is
\[
k_{BH}=(\bar{p}_R+\bar{p}_L,\bar{p}_R-\bar{p}_L)\;.
\]
The black hole thus moves with respect to the lab frame with velocity
\[
v={\bar{p}_R-\bar{p}_L\over \bar{p}_R+\bar{p}_L}\;.
\]
The time delays measured by the detectors are related to the evaporation time as measured in the lab frame by
\[
\Delta t_{lab}=\frac{\Delta t_{L,R}}{1\pm v}=\frac12\ell_s^2(\bar{p}_R+\bar{p}_L)\,,
\]
where the upper/lower sign should be used for the time delay measured by left/right detector. The evaporation time in the rest frame of the black hole is then simply
\[\Delta t_{cms}=\ell_s^2\sqrt{\bar{p}_L\bar{p}_R}=\frac12\ell_s^2 E_{cms}\,,\]
confirming that the evaporation time is simply linear in the mass consistent with the black hole interpretation.

To make the case for the black hole interpretation of the $S$-matrix (\ref{S-matrix}) even stronger, note that there is an equivalence principle at work. Let us replace the right-moving particle with a collection of $N$ soft particles, each carrying a $1/N$ fraction of the total energy.
Then the factorizability of the $S$-matrix implies that the total time delay is 
\be
\label{1/Nshift}
\Delta t_{N}=N2\delta_{NG}(4\bar{p}_R\bar{p}_L)\bar{p}^{-1}_L\;.
\ee
Linearity of the Nambu--Goto phase shift $\delta_{NG}(s)$ implies that the two time delays (\ref{1/Nshift}) and (\ref{evaporation_time}) are the same. In other words, as expected in a gravitational theory, the overall phase shift is determined by the total energy of the collision alone (cf. (\ref{psiout})).

To conclude this section, let us point out an interesting similarity between the integrals encountered here, such as (\ref{out-state}), and those appearing in the Moyal product in non-commutative theories, with the string length $\ell_s^2$ being similar to the non-commutativity parameter $\theta$. Indeed, some of the properties of the scattering amplitude discussed here were also observed in the tree level scattering in field theories with space/time non-commutativity \cite{Seiberg:2000gc}. That amplitude, however, contains an additional constant contribution as well as an unphysical $e^{-is\ell_s^2}$ term leading to acausal behavior. This is a good illustration that the exponential boundedness on the physical sheet, enjoyed by the amplitude (\ref{S-matrix}), is crucial for a consistent space-time interpretation. In our case it is the configuration space for multi-particle states that seems to become non-commutative as indicated by equation~\eqref{eq:su} rather than space-time. 

\section{Classical Solutions: Black Hole Precursors and Cosmology}
\label{sec:classical}
Typically, a discussion of a gravitational theory starts with the analysis of its classical solutions. We were lucky and were able to present the exact quantum solution first, so that the classical dynamics had to wait until the very end. Nevertheless, it is instructive to discuss it briefly, to illustrate that the worldsheet theory exhibits features characteristic for a gravitational theory even at the classical level, again modulo limitations related to integrability.
It is well-known how to describe general classical solutions of the Nambu--Goto action \cite{Goddard:1973qh}, and we will not give a general overview.
Instead we choose to describe two classes of solutions, which represent the footprints of gravitational physics, black holes (or, more accurately, integrable black hole precursors), and cosmologies.
\subsection{Classical origin of time delay}
One simple fact about classical string dynamics to keep in mind is that the causal structure on the world-sheet is determined by the induced metric
\be
\label{induced}
h_{\alpha\beta}=\d_\alpha X^\mu\d_\beta X_\mu\;.
\ee
We will present the solutions in static gauge $X^0=\tau$, $X^1=\sigma$, because we feel that in this gauge the meaning and the origin of the $S$-matrix (\ref{S-matrix}) are the most transparent. 

From our experience with conventional Einstein gravity, one expects that for large center of mass energies $s\ell_s^2\gg 1$ the origin of the time delay (\ref{evaporation_time}) should be seen  semiclassicaly. The corresponding time delay in four-dimensional gravity follows from classical black hole formation and its consequent evaporation, as described by the Hawking calculation, which represents the leading quantum (one-loop) correction to the classical solution.

Note that any purely left-moving configuration $X_{cl}^i(\tau+\sigma)$ presents 
an exact solution of the Nambu--Goto equations. A natural string configuration to study the (semi)classical origin of the time delay (\ref{evaporation_time}) is then a large left-moving kink $X_{cl}^i(\tau+\sigma)$
acting as a background. For simplicity, we consider a configuration with a single non-zero flavor $X^i_{cl}(\tau+\sigma)$, and suppress the flavor superscript in what follows. 

One then sends a small right-moving perturbation across this kink, and calculates the time it takes to reach the other side.
In the probe approximation this amounts to studying right-moving null geodesics in the induced metric (\ref{induced}), corresponding to $X_{cl}(\tau+\sigma)$
\be
\label{acoustic_metric}
ds^2=(-1+X'^2_{cl})d\tau^2+2X'^2_{cl}d\tau d\sigma+(1+X'^2_{cl})d\sigma^2\,.
\ee
The null geodesic equation results in
\be
\label{geodesic}
\dot{\sigma}(\tau)={-X'^2_{cl}\l\tau+\sigma(\tau)\r\pm 1\over X'^2_{cl}(\tau+\sigma(\tau))+1}\;.
\ee
The lower sign corresponds to a left-mover, whose propagation is unaffected by the presence of the background. The upper sign corresponds to a right-mover,
which experiences a time delay
\be
\label{classical_delay}
\Delta t=\int_{-\infty}^{\infty}d\tau\l1-\dot{\sigma}\r =\int_{-\infty}^{\infty}d\tau{2X'^2_{cl}(\tau+\sigma_0(\tau))\over  X'^2_{cl}(\tau+\sigma_0(\tau))+1}=\int_{-\infty}^{\infty}dzX'^2_{cl}(z)\;.
\ee
Here $\sigma_0(\tau)$ is a solution of the geodesic equation (\ref{geodesic}), and we used equation~(\ref{geodesic}) when changing the integration variable from $\tau$ to $z=\tau+\sigma_0(\tau)$. We recognize
\[
\int_{-\infty}^{\infty}dzX'^2_{cl}(z)=\ell_s^2\Delta E\;,
\]
where $\Delta E$ is the energy of the classical solution $X_{cl}$ relative to the vacuum energy. The time delay obtained in the classical theory~(\ref{classical_delay}) thus exactly coincides with the one derived in the quantum theory~(\ref{evaporation_time}). This is an important difference with a realistic quantum theory of gravity -- as a consequence of integrability there is no actual horizon and particle production and the black hole ``evaporation" happens classically. 

\begin{figure}[t!] 
 \begin{center}
 \includegraphics[width=3.5in]{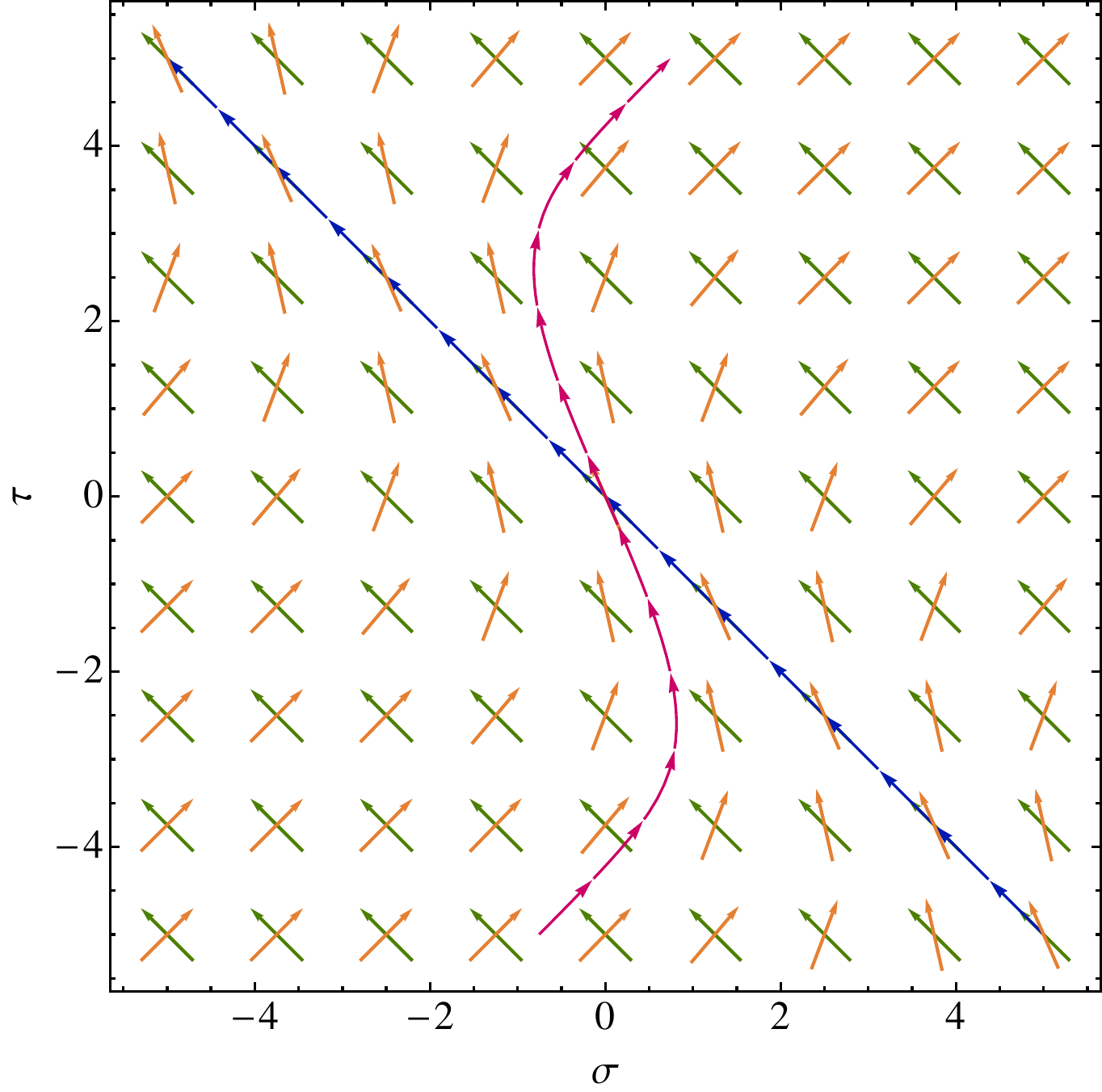}
 \caption{This figure shows the structure of light-cones indicated by the green and orange arrows in the background of a left-moving kink in the static gauge. The blue and red lines show geodesics of a left- and right-moving particle, respectively.}
 \label{null_vectors}
 \end{center}
\end{figure}

In spite of this deficiency, we feel that this class of solutions is close in other respects to actual black holes. For instance, it follows from (\ref{geodesic}) that for $X'^2_{cl}>1$ there is a region inside the kink where both left- and right-movers propagate towards the left. We see that the classical origin of the time delay (\ref{evaporation_time}) is very intuitive---a right-mover gets carried away towards left by the kink, see Figure~\ref{null_vectors}.
In the presence of non-integrable perturbations, when the energy transfer between left and right-movers becomes possible, one expects the emergence of an actual horizon.

The CGHS model \cite{Callan:1992rs} supports this expectation. The field equations, with quantum backreaction taken into account, look a lot like
the equations of the bosonic string in the Polyakov formalism with additional interactions included. Purely left-moving excitations of matter fields (which are analogous to our $X^i$)  solve the field equations, but do give rise to a black hole horizon. It appears plausible that the $S$-matrix (\ref{S-matrix}) provides a reasonable zeroth order approximation for the process of black hole evaporation in models of this type.

Note, that for a right-moving bump excited in the same target space direction as the left-moving one (and perhaps in one other transverse direction) the classical string solutions generically develop cusp singularities. These singularities are not shielded by any horizon. The development  of  classical singularities is another consequence of the absence of  power counting renormalizability.
In power counting renormalizable theories  singularities usually do not occur if one starts with a regular initial data \cite{Eardley:1982dn,Eardley:1982dm}. They do occur for the string worldsheet theory but get resolved at the quantum level by the $S$-matrix (\ref{S-matrix}).

At the end of section~\ref{sec:loop}, we mentioned that our $S$-matrix is related to a trivial $S$-matrix by a gauge transformation which acts non-trivially at infinity. We are now in a position to provide a physical interpretation for one natural choice of coordinates related to the static gauge coordinates by a gauge transformation of this kind. To this end, it is helpful to write the metric~\eqref{acoustic_metric} in a form that makes its null Killing vector manifest
\be
ds^2=-d\tau^+d\tau^-+X_{cl}'^2(\tau^+)(d\tau^+)^2\,,
\ee 
where $\tau^\pm=\tau\pm\sigma$. We then see that in coordinates $t,x$ related to $\tau^\pm$ by
\be\label{eq:LCcoordtrans}
\tau^+=t+x\qquad\text{and}\qquad\tau^-=t-x+\int_{-\infty}^{t+x}\,dz\,X_{cl}'^2(z)\,,
\ee
the metric is simply the Minkowski metric
\be\label{eq:LCMink}
ds^2=-dt^2+dx^2\,.
\ee
This means that the coordinates $(t,x)$ correspond to a family of free-falling timelike observers. The coordinate time $t$ is the proper time along their worldlines and the worldlines of these observers are labeled by their positions $x$ in target space long before the kink approaches them.\footnote{To see this, note that the geodesics behave like $X^0(t)\sim t$ and $X^1(t)\sim x$ as $t\to-\infty$.}
In these coordinates, a right-moving probe moves along a worldline $t=x+const$ whether a kink exists or not. So a free-falling detector will not measure a time-delay, unlike a detector that is mounted at some constant value $X^1$ which measures the time delay~\eqref{classical_delay}.

The classical time delay~\eqref{classical_delay} is closely tied to the $S$-matrix~\eqref{S-matrix} and the observation that freely falling observers do not measure a time delay suggests that they observe a trivial $S$-matrix. However, there is still one small step needed to show that this is indeed the case. We have to extend our discussion to configurations that contain left- and right- moving excitations of comparable energies. By definition of the $S$-matrix the detectors should be far away from the interaction region so that, at any given time, they may be close to either the left- or the right-moving excitation, but not both. The effects of left- and right-moving excitations thus simply add and the change of coordinates   
\be\label{eq:LCcoords}
\tau^+=t+x+\int_{-\infty}^{t-x}\,dz\,(\partial_-X^i)^2(z)\qquad\text{and}\qquad\tau^-=t-x+\int_{-\infty}^{t+x}\,dz\,(\partial_+X^i)^2(z)\,,
\ee
leads to a metric that, for Gaussian wave packets, is exponentially well approximated by the flat Minkowski metric~\eqref{eq:LCMink} far away from the interaction region, and a metric that is conformally related to it where right- and left- movers are not well separated. With this choice of coordinates the constraints are solved classically as in light-cone quantization and one can show that the fields corresponding to the transverse fluctuations enjoy the mode expansions of free fields. The corresponding trivial $S$-matrix, seen by putative free-falling detectors on the worldsheet, is often referred to as the worldsheet $S$-matrix. Our $S$-matrix is observed by detectors freely falling with respect to the target space metric. It is usually called the target space $S$-matrix, in this case for scattering of worldsheet degrees of freedom. 
The two $S$-matrices are related by a gauge transformation~\eqref{eq:LCcoords}, but are inequivalent because this gauge transformation acts non-trivially at infinity. More specifically, as \mbox{$t\to-\infty$}, the coordinates $t$ and $x$ agree with the target space coordinate $\tau$ and $\sigma$ while at late times they differ by a constant shift proportional to the energy and momentum of the configuration.

In this case, we can also be more specific about precisely how the $X^i(t,x)$ fail to satisfy~\eqref{eq:lt} for the Poincar\'e transformations compatible with our $S$-matrix. In writing~\eqref{eq:LCcoords}, we have singled out one particular family of inertial observers. They are initially at rest with respect to the target space and at early times their coordinate system and the target space coordinate system agree. To meaningfully talk about a worldsheet scattering, we should not gauge-fix the global Poincar\'e group and we should introduce zero-modes $x^\pm$, $p^\pm$ corresponding to translations and boosts relative to the target space into the fields $X^\pm(t,x)$ which modulo these zero modes are given by~\eqref{eq:LCcoords}. The Poincar\'e group that is compatible with our $S$-matrix then acts only on these zero modes, leaving the $X^i(t,x)$ invariant, incompatible with~\eqref{eq:lt}. They do, of course, transform appropriately under the Poincar\'e group associated with the worldsheet $S$-matrix. 

Returning to the black hole analogy, the two different $S$-matrices just described are measured by an infalling and an outside observer. Somewhat surprisingly, in this simple theory an infalling observer finds a free $S$-matrix and is able to construct local observables, which are simply the free correlators of $X^i$'s in the light cone gauge, they behave non-locally from the viewpoint of an outside observer. Their existence is contrary to the experience in realistic theories of gravity, where the construction of sharply defined observables  for an infalling observer seems problematic. This may be an artifact of integrability.
\subsection{Cosmology}
Let us conclude our discussion of classical world-sheet solutions by presenting a solution with properties characteristic of cosmologies.
For now we continue to work in static gauge, and for simplicity  again consider a single non-vanishing flavor $X$. Ordinarily, cosmological solutions are understood to be homogeneous (isotropy is not a restriction in a $(1+1)$-dimensional world).
However, the only $\sigma$-independent solutions of the Nambu--Goto theory are those linear in $\tau$, which is equivalent to the trivial vacuum solution. To find a non-trivial solution possessing a high degree of symmetry, let us instead consider boost invariant string configurations of the form
$
X(\sigma^2-\tau^2).
$
The solutions to the Nambu--Goto field equations with this Ansatz are easily found, but it turns out that static gauge is not very convenient to understand the global structure of the solution. So let us present the solution as a hypersurface in the target space-time.
It consists of two branches $A,B$
\bea
&&{X_A^{0}}^{2}-{X_A^{1}}^{2}-L_0^2(\sinh (X_A/L_0))^2=0\,,\\
&&{X_B^{1}}^{2}-{X_B^{0}}^{2}-L_0^2(\cosh (X_B/L_0))^2=0\,.
\eea
Here $L_0$ is a characteristic length scale. We work in units such that $L_0=1$ in what follows to keep the formulas shorter.
To understand the physics of these solutions, it is convenient  to choose the boost parameter as one of the worldsheet coordinates, which  makes the corresponding isometry manifest. Thus, for the $A$-branch we write,
\[
X_A^0=\rho \cosh\lambda\;,\;\;X_A^1=\rho\sinh\lambda\;.
\]
Then the induced metric takes the following form
\be
\label{FRW}
ds_A^2=-{\rho^2\over 1+\rho^2}d\rho^2+\rho^2 d\lambda^2=-d\tau^2+(\tau^2+2\tau)d\lambda^2\;,
\ee
where we changed coordinates from $\rho$ to $\tau=\sqrt{1+\rho^2}-1$. We see that this branch corresponds to an
expanding ($\rho>0$) and contracting ($\rho<0$) cosmological solution. It approaches the flat Milne universe at late and early times and passes through a  Big Bang/Crunch singularity at $\rho=0$, or in target-space coordinates at $X_A^0=\pm X_A^1$.   Close to the singularity the expansion rate is the same as in a $4$d radiation dominated Universe.

Following the same logic for the $B$-branch, we choose
\[
X_B^0=\rho \sinh\lambda\;,\;\;X_B^1=\rho\cosh\lambda\;.
\]
Note, that this branch actually corresponds to two identical disconnected surfaces, corresponding to $\rho\geq 1$ and $\rho\leq -1$.
Choosing the former, we obtain the  static induced metric of the form
\be
\label{Bbranch}
ds^2=-\rho^2d\lambda^2+{\rho^2\over \rho^2-1}d\rho^2=-(1+r^2)d\lambda^2+dr^2\;,
\ee
where $r=\sqrt{\rho^2-1}$. Note that for each pair $(\lambda,\rho)$ there are two points on the solution with different sign of $X$. This can be accounted for by extending the range of $r$ in (\ref{Bbranch}) to $r\in(-\infty,\infty)$. This static solution is asymptotically flat at large $r$, with $\lambda,r$ turning into the Rindler coordinates.

\section{Conclusions, Speculations and Future Directions}
\label{conclusions}
We hope to have convinced the reader that the worldsheet theory of an infinitely long ``free" bosonic string is non-trivial and comes as close to capturing essential features of gravitational dynamics as one can hope for from a two-dimensional integrable theory.
Let us conclude by outlining several directions in which one can further elaborate on this observation. Somewhat loosely these can be divided into understanding more about the dynamics of gravitational theories in two dimensions and into extracting general lessons about gravitational theories in other numbers of dimensions.

The first natural ``two-dimensional" question is how to extract information about the non-linearly realized target space Poincar\'e symmetry directly from our exact $S$-matrix. In particular, it should be rather satisfying to see $D=26$ arise in this language.

We have only discussed the bosonic string so far. One may wonder how all this extends to worldsheet theories of supersymmetric strings. The matter content is then extended by fermions as required by supersymmetry. From the known light-cone spectra of superstring theories one finds once again that the $S$-matrix reduces to the diagonal phase shift~(\ref{S-matrix}). Understanding how the non-linearly realized super-Poincar\'e symmetry can be extracted from this $S$-matrix alone would certainly also be rather gratifying.

As soon as one extends the discussion to superstrings, it is natural to ask what other options exist for constructing integrable gravitational theories. An example of another gravitational $S$-matrix that still enjoys analyticity, unitarity, crossing symmetry and exponential boundedness on the physical sheet would be
\be
\label{gravgoldstino}
\e^{2i\delta(s)}={iM^2-s\over iM^2+s}\e^ {i s\ell_s^2/4}\;.
\ee
It is natural to suggest, that at least for $\ell_s M\ll1$, this $S$-matrix describes the goldstino theory coupled to gravity, similarly to how the $S$-matrix (\ref{S-matrix})
may be thought of as describing $(D-2)$ free bosons coupled to gravity.
The latter interpretation can be made  more manifest by noting that at the classical level the theory with the light-cone spectrum~\eqref{LCspectrum} can be formulated in the Polyakov formalism
as
\be
\label{Polyakov}
S_{D-2}=\int d^2\sigma\sqrt{-g}\l g^{\alpha\beta}(-\d_\alpha X^0\d_\beta X^0+\d_\alpha X^1\d_\beta X^1)+{\cal L}_{D-2}(g,X^i)\r\;.
\ee
where ${\cal L}_{D-2}(g,X^i)$ describes $(D-2)$ free bosons coupled to the metric. The conventional approach to quantizing this theory uses (\ref{Polyakov}) as the starting point, resulting in Liouville gravity in the non-critical case.
Instead, the exact $S$-matrix (\ref{S-matrix}) results from   solving for  the auxiliary metric $g$ at the classical level with subsequent quantization in light-cone gauge. The two procedures are not equivalent. For instance, the latter does not preserve the full $ISO(D-1,1)$ symmetry away from the critical dimension.
It is possible to generalize this procedure by replacing  ${\cal L}_{D-2}(g,X^i)$ with a general conformal field theory ${\cal L}_{CFT}$ and use this as a working definition of what one means by coupling the theory to gravity.  For non-conformal theories (as needed to obtain (\ref{gravgoldstino})), this proposal cannot literally work because there the Liouville mode already appears at the classical level.

It is also natural to ask whether integrable massive gravitational theories ({\it i.e.}, exhibiting $\e^{is \ell_s^2/4}$-type  non-analyticities in the UV) can be constructed.
At the $S$-matrix level the massive generalization of the amplitude (\ref{S-matrix}) is
\be
\label{massive}
e^{2i\delta(s)}=e^{i\ell_s^2\sqrt{s^2-4m^2 s}/4}
\ee
It would be interesting to work out what the Lagrangian description of such a theory is.

Furthermore, it is interesting to study whether physically sound integrable gravitational theories with non-diagonal scattering exist.
It is straightforward to check that non-trivial annihilations in a massless $SO(D-2)$-invariant theory contradict the Yang--Baxter equation in the absence of left-left and right-right scattering. This rules out the possibility for integrable non-critical $ISO(D-1,1)$-invariant effective string theories away from the critical dimension, because the Polchinski--Strominger interaction \cite{Polchinski:1991ax} introduces annihilations~\cite{effective}.
However, this still leaves  room for a large number of other possibilities.

Turning to the questions of interest beyond the two-dimensional world it seems interesting to understand to what extent knowledge of the exact flat space $S$-matrix (and finite volume spectrum) sheds light on how and whether the ``Big Bang" singularity described in section~\ref{sec:classical} is resolved.

Another interesting set of questions comes from thinking about how the full interacting string theory fits in the picture.
As we already mentioned at the end of section~\ref{sec:loop}, it manifests itself in the existence of additional non-local off-shell observables---perturbative string amplitudes. Their existence indicates that the worldsheet theory---a UV complete two-dimensional theory of gravity on its own---can be embedded in a larger and richer structure, the interacting string theory. 

This is similar to how the existence of off-shell observables---local correlation functions---in conventional field theories indicate that they can be extended to larger field theories, incorporating dynamical probes coupled to the corresponding operators. The difference in the present case appears to be that the off-shell observables are non-local. 
One may wonder, whether this process stops at the level of string/M-theory, or whether it, too, can be a part of an even larger structure.

Note that the embedding into a larger structure typically allows the definition of a larger set of physical observables.
For example, $D$-branes provide space-time probes at substringy scales \cite{Douglas:1996yp}.
 It may be that difficulties with resolving space-like singularities  within string theory hint for the existence of such a structure, which would provide a theory for Big Bang initial conditions, for example. Perhaps,
a better understanding of the cosmological solution of section~\ref{sec:classical} can shed some light on this.

This viewpoint is somewhat similar to the approach of \cite{Banks:1988je,Hawking:1990ue,Rubakov:1995zb} where the worldsheet theory was suggested as a toy model to the study of Euclidean wormhole processes, which correspond to string splittings.
In light of our arguments this analogy appears even more appropriate, given that the ``free" worldsheet theory itself can be considered as a non-trivial gravitational theory.

 Interestingly, it seems impossible to decide based on the knowledge of the free worldsheet theory alone whether two-dimensional wormholes (aka string splitting processes) are physical or not. These are unphysical if $g_s=0$ and physical otherwise. This casts doubts on the argument~\cite{ArkaniHamed:2007js} that higher dimensional wormholes should be unphysical saddle points, based on the observation that Euclidean wormholes exist in the bulk in certain AdS/CFT setups while the CFT knows nothing about them. 
Just like our $S$-matrix~(\ref{S-matrix})  shows no signs of the existence of string splittings, but can be extended to allow for these, it might be that the CFT can be extended to make the wormhole processes physical.
 
 As a final remark, note that the setup considered here is reminiscent of ``asymptotic safety"~\cite{Weinberg:safety}---a naively non-renormalizable theory 
turns out to be UV complete and provides a set of well-defined physical observables. However, the situation appears to be more delicate. A classic example of asymptotic safety is provided by the goldstino $S$-matrix (\ref{goldstino}). Here indeed, in perfect agreement with the asymptotic safety proposal,
a naively non-renormalizable theory turns out to have a non-trivial UV fixed point, and gives rise to a perfectly conventional, even if strongly coupled RG flow. What happens in the worldsheet theory is rather different and is perhaps more appropriately called ``asymptotic fragility". The theory does provide a well-defined $S$-matrix (\ref{S-matrix}). However, 
there is  strong evidence that  no UV fixed point exists, and  attempts to introduce local probes destroy the finiteness.
\section*{Acknowledgments}
\addcontentsline{toc}{section}{Acknowledgments}
We thank Nima Arkani-Hamed, Giga Gabadadze,  Matt Kleban, Zohar Komargodski, Juan Maldacena, Mehrdad  Mirbabayi, Alberto Nicolis, Matt Roberts, Slava Rychkov, Peter Tinyakov, Arkady Tseytlin, Giovanni Villadoro, and Ed Witten for useful discussions.
This work is supported in part by the NSF grant PHY-1068438. The work of R.F. has been supported in part by the National Science Foundation under Grant No. NSF-PHY-0855425. R.F. would like to thank the Aspen Center for Physics for hospitality during the final stages of this work.
\appendix
\section*{Appendix: Calculation of the Entanglement Entropy}
\addcontentsline{toc}{section}{Appendix: Calculation of the Entanglement Entropy}
\label{appendix}
Let us calculate the entanglement entropy corresponding to the density matrix (\ref{scrambled}).
We will use the expression
\[
S_{ent}\equiv-\Tr \rho\log\rho=-\left.{\d\over\d\gamma}\Tr\rho^\gamma\right|_{\gamma=1}\;.
\]
For  $\Tr\rho^\gamma$ we will use the analytic continuation from integer values of $\gamma$. With the density matrix (\ref{momentumrho}) we have
\be
\label{Trn}
\Tr\rho^n=\int \prod_{i=1}^n dp_{Ri}dp_{Li}|f(p_{Ri})|^2|f(p_{Li})|^2e^{ip_{Ri}(p_{Li}-p_{L(i+1)})\ell_s^2}\;,
\ee 
where $p_{L(n+1)}\equiv p_{L1}$ and we extended the momentum integration range over the whole real axis, assuming that the wave packets are sufficiently narrow in the momentum. We consider the case of Gaussian packets,
\[
|f(p_{L(R)})|^2=
{1\over (2\pi)^{1/2}\Delta p_{L(R)}}
e^{-{
(p_{L(R)}-
\bar{ p}_{L(R)})^2\over 2\Delta p_{L(R)}^2
            }}\;,
\]
so that (\ref{Trn}) takes the form
\be
\label{TrnGauss}
\Tr\rho^n=\int \prod_{i=1}^n{ dp_{Ri}dp_{Li}\over 2\pi \Delta p_L \Delta p_R}e^{ip_{Ri}(p_{Li}-p_{L(i+1)})\ell_s^2-{p_{Li}^2\over 2\Delta p_{L}^2}-{p_{Ri}^2\over 2\Delta p_{R}^2}}\;.
\ee 
Note, that in the limit $\ell_s\to 0$ this integral is proportional to the periodic Euclidean path integral for the harmonic oscillator in the canonical phase space labeled by $(p_R,p_L)$. It will be interesting to find a physical interpretation for this result.
By performing a Gaussian integral over the ``canonical momenta"
$p_{Ri}$ we obtain the path integral in the coordinate form,
 \be
\label{TrnGaussLagr}
\Tr\rho^n=\int \prod_{i=1}^n{ dp_{Li}\over (2\pi)^{1/2} \Delta p_L}e^{-{(p_{Li}-p_{L(i+1)})^2\ell_s^4 \Delta p_{R}^2\over 2}-{p_{Li}^2\over 2\Delta p_{L}^2}}\;.
\ee 
In the limit $\ell_s^2\Delta p_R\Delta p_L\gg1$ and $n\gg 1$ this integral can be rewritten as
\be
\label{rhonosc}
\Tr\rho^n={Z(\beta)\over \l \ell_s^2\Delta p_R\Delta p_L\r^n}\;,
\ee
where 
\[
Z(\beta)={e^{-\beta/2}\over 1-e^{-\beta}}
\] 
is a thermal partition function for a harmonic oscillator of a unit frequency at the inverse temperature equal to
\[
\beta={n\over \ell_s^2\Delta p_L\Delta p_R}\;.
\]
Using (\ref{rhonosc}) as a starting point for the analytic continuation to real $n$ we obtain the following estimate for the entanglement entropy,
\be
\label{entropy}
S_{ent}=\log \ell_s^2\Delta p_L\Delta p_R\;.
\ee 
Note, that (\ref{TrnGaussLagr}) can be computed exactly, giving
\be
\label{rhonex}
\Tr\rho^n= \l -2 \alpha^n+{1\over 2^n}\l \l 1+2\alpha-\sqrt{1+4\alpha} \r^n + \l 1+2\alpha+\sqrt{1+4\alpha} \r^n \r\r^{-1/2},
\ee
Where $\alpha=\ell_s^4\Delta p_R^2\Delta p_L^2$. The leading large $\alpha$ behavior of the entropy resulting from this expression, is the same as (\ref{entropy}). 
\bibliographystyle{utphys}
\bibliography{dlrrefs}

\providecommand{\href}[2]{#2}\begingroup\raggedright\begin{thebibliography}{10}

\bibitem{Maldacena:1997re}
J.~M. Maldacena, ``{The Large N limit of superconformal field theories and
  supergravity},'' {\em Adv.Theor.Math.Phys.} {\bf 2} (1998) 231--252,
\href{http://www.arXiv.org/abs/hep-th/9711200}{{\tt hep-th/9711200}}.

\bibitem{Gubser:1998bc}
S.~Gubser, I.~R. Klebanov, and A.~M. Polyakov, ``{Gauge theory correlators from
  noncritical string theory},'' {\em Phys.Lett.} {\bf B428} (1998) 105--114,
\href{http://www.arXiv.org/abs/hep-th/9802109}{{\tt hep-th/9802109}}.

\bibitem{Witten:1998qj}
E.~Witten, ``{Anti-de Sitter space and holography},'' {\em
  Adv.Theor.Math.Phys.} {\bf 2} (1998) 253--291,
\href{http://www.arXiv.org/abs/hep-th/9802150}{{\tt hep-th/9802150}}.

\bibitem{Witten:1995ex}
E.~Witten, ``{String theory dynamics in various dimensions},'' {\em Nucl.Phys.}
  {\bf B443} (1995) 85--126,
\href{http://www.arXiv.org/abs/hep-th/9503124}{{\tt hep-th/9503124}}.

\bibitem{Banks:1996vh}
T.~Banks, W.~Fischler, S.~Shenker, and L.~Susskind, ``{M theory as a matrix
  model: A Conjecture},'' {\em Phys.Rev.} {\bf D55} (1997) 5112--5128,
\href{http://www.arXiv.org/abs/hep-th/9610043}{{\tt hep-th/9610043}}.

\bibitem{Luscher:1986pf}
M.~Luscher, ``{Volume Dependence of the Energy Spectrum in Massive Quantum
  Field Theories. 2. Scattering States},'' {\em Commun.Math.Phys.} {\bf 105}
  (1986)
153--188.

\bibitem{Luscher:1990ck}
M.~Luscher and U.~Wolff, ``{How To Calculate The Elastic Scattering Matrix In
  Two-Dimensional Quantum Field Theories By Numerical Simulation},'' {\em
  Nucl.Phys.} {\bf B339} (1990)
222--252.

\bibitem{ArkaniHamed:2007ky}
N.~Arkani-Hamed, S.~Dubovsky, A.~Nicolis, E.~Trincherini, and G.~Villadoro,
  ``{A Measure of de Sitter entropy and eternal inflation},'' {\em JHEP} {\bf
  0705} (2007) 055,
\href{http://www.arXiv.org/abs/0704.1814}{{\tt 0704.1814}}.

\bibitem{Weinberg:1980kq}
S.~Weinberg and E.~Witten, ``{Limits on Massless Particles},'' {\em Phys.Lett.}
  {\bf B96} (1980)
59.

\bibitem{Cardy:1984rp}
J.~L. Cardy, ``{Conformal invariance and universality in finite-size
  scaling},'' {\em J.Phys.A} {\bf 17} (1984)
L385--L387.

\bibitem{Zamolodchikov:1989cf}
A.~Zamolodchikov, ``{Thermodynamic Bethe Ansatz In Relativistic Models. Scaling
  Three State Potts And Lee-Yang Models},'' {\em Nucl.Phys.} {\bf B342} (1990)
695--720.

\bibitem{Callan:1992rs}
C.~G. Callan, S.~B. Giddings, J.~A. Harvey, and A.~Strominger, ``{Evanescent
  black holes},'' {\em Phys.Rev.} {\bf D45} (1992) 1005--1009,
\href{http://www.arXiv.org/abs/hep-th/9111056}{{\tt hep-th/9111056}}.

\bibitem{Russo:1992ht}
J.~G. Russo, L.~Susskind, and L.~Thorlacius, ``{Black hole evaporation in
  (1+1)-dimensions},'' {\em Phys.Lett.} {\bf B292} (1992) 13--18,
\href{http://www.arXiv.org/abs/hep-th/9201074}{{\tt hep-th/9201074}}.

\bibitem{Fiola:1994ir}
T.~M. Fiola, J.~Preskill, A.~Strominger, and S.~P. Trivedi, ``{Black hole
  thermodynamics and information loss in two-dimensions},'' {\em Phys.Rev.}
  {\bf D50} (1994) 3987--4014,
\href{http://www.arXiv.org/abs/hep-th/9403137}{{\tt hep-th/9403137}}.

\bibitem{Ashtekar:2010hx}
A.~Ashtekar, F.~Pretorius, and F.~M. Ramazanoglu, ``{Surprises in the
  Evaporation of 2-Dimensional Black Holes},'' {\em Phys.Rev.Lett.} {\bf 106}
  (2011) 161303,
\href{http://www.arXiv.org/abs/1011.6442}{{\tt 1011.6442}}.

\bibitem{Hawking:1976ra}
S.~Hawking, ``{Breakdown of Predictability in Gravitational Collapse},'' {\em
  Phys.Rev.} {\bf D14} (1976)
2460--2473.

\bibitem{Ginsparg:1993is}
P.~H. Ginsparg and G.~W. Moore, ``{Lectures on 2-D gravity and 2-D string
  theory},''
\href{http://www.arXiv.org/abs/hep-th/9304011}{{\tt hep-th/9304011}}.

\bibitem{effective}
S.~Dubovsky, R.~Flauger, and V.~Gorbenko, ``{Effective String Theory
  Revisited},''
\href{http://www.arXiv.org/abs/1203.1054}{{\tt 1203.1054}}.

\bibitem{Polchinski:1991ax}
J.~Polchinski and A.~Strominger, ``{Effective string theory},'' {\em
  Phys.Rev.Lett.} {\bf 67} (1991) 1681--1684.

\bibitem{Polchinski:1998rq}
J.~Polchinski, ``{String theory. Vol. 1: An introduction to the bosonic
  string},''.

\bibitem{Fendley:1993jh}
P.~Fendley and H.~Saleur, ``{Massless integrable quantum field theories and
  massless scattering in (1+1)-dimensions},''
\href{http://www.arXiv.org/abs/hep-th/9310058}{{\tt hep-th/9310058}}.

\bibitem{Zamolodchikov:1991vx}
A.~Zamolodchikov, ``{From tricritical Ising to critical Ising by thermodynamic
  Bethe ansatz},'' {\em Nucl.Phys.} {\bf B358} (1991)
524--546.

\bibitem{Klassen:1989ui}
T.~R. Klassen and E.~Melzer, ``{Purely Elastic Scattering Theories and their
  Ultraviolet Limits},'' {\em Nucl.Phys.} {\bf B338} (1990)
485--528.

\bibitem{ArkaniHamed:2008gz}
N.~Arkani-Hamed, F.~Cachazo, and J.~Kaplan, ``{What is the Simplest Quantum
  Field Theory?},'' {\em JHEP} {\bf 1009} (2010) 016,
\href{http://www.arXiv.org/abs/0808.1446}{{\tt 0808.1446}}.

\bibitem{Kastor:1988ef}
D.~Kastor, E.~Martinec, and S.~Shenker, ``{RG Flow in N=1 Discrete Series},''
  {\em Nucl.Phys.} {\bf B316} (1989)
590--608.

\bibitem{Dorey:1996re}
P.~Dorey and R.~Tateo, ``{Excited states by analytic continuation of TBA
  equations},'' {\em Nucl.Phys.} {\bf B482} (1996) 639--659,
\href{http://www.arXiv.org/abs/hep-th/9607167}{{\tt hep-th/9607167}}.

\bibitem{Teschner:2007ng}
J.~Teschner, ``{On the spectrum of the Sinh-Gordon model in finite volume},''
  {\em Nucl.Phys.} {\bf B799} (2008) 403--429,
\href{http://www.arXiv.org/abs/hep-th/0702214}{{\tt hep-th/0702214}}.

\bibitem{Smirnov:1992vz}
F.~Smirnov, ``{Form-factors in completely integrable models of quantum field
  theory},'' {\em Adv.Ser.Math.Phys.} {\bf 14} (1992)
1--208.

\bibitem{Delfino:1994ea}
G.~Delfino, G.~Mussardo, and P.~Simonetti, ``{Correlation functions along a
  massless flow},'' {\em Phys.Rev.} {\bf D51} (1995) 6620--6624,
\href{http://www.arXiv.org/abs/hep-th/9410117}{{\tt hep-th/9410117}}.

\bibitem{Seiberg:2000gc}
N.~Seiberg, L.~Susskind, and N.~Toumbas, ``{Space-time noncommutativity and
  causality},'' {\em JHEP} {\bf 0006} (2000) 044,
\href{http://www.arXiv.org/abs/hep-th/0005015}{{\tt hep-th/0005015}}.

\bibitem{Goddard:1973qh}
P.~Goddard, J.~Goldstone, C.~Rebbi, and C.~B. Thorn, ``{Quantum dynamics of a
  massless relativistic string},'' {\em Nucl.Phys.} {\bf B56} (1973)
109--135.

\bibitem{Eardley:1982dn}
D.~Eardley and V.~Moncrief, ``{The Global Existence Of Yang-Mills Higgs Fields
  In Four-Dimensional Minkowski Space. 2. Completion Of Proof},'' {\em
  Commun.Math.Phys.} {\bf 83} (1982)
193--212.

\bibitem{Eardley:1982dm}
D.~Eardley and V.~Moncrief, ``{The Global Existence Of Yang-Mills Higgs Fields
  In Four-Dimensional Minkowski Space. 1. Local Existence And Smoothness
  Properties},'' {\em Commun.Math.Phys.} {\bf 83} (1982)
171--191.

\bibitem{Douglas:1996yp}
M.~R. Douglas, D.~N. Kabat, P.~Pouliot, and S.~H. Shenker, ``{D-branes and
  short distances in string theory},'' {\em Nucl.Phys.} {\bf B485} (1997)
  85--127,
\href{http://www.arXiv.org/abs/hep-th/9608024}{{\tt hep-th/9608024}}.

\bibitem{Banks:1988je}
T.~Banks, ``{Prolegomena to a Theory of Bifurcating Universes: A Nonlocal
  Solution to the Cosmological Constant Problem Or Little Lambda Goes Back to
  the Future},'' {\em Nucl. Phys.} {\bf B309} (1988)
493.

\bibitem{Hawking:1990ue}
S.~W. Hawking, ``{The Effective action for wormholes},'' {\em Nucl. Phys.} {\bf
  B363} (1991)
117--131.

\bibitem{Rubakov:1995zb}
V.~Rubakov, ``{Modeling macroscopic and baby universes by fundamental
  strings},'' {\em Nucl.Phys.} {\bf B453} (1995) 395--412,
\href{http://www.arXiv.org/abs/hep-th/9505159}{{\tt hep-th/9505159}}.

\bibitem{ArkaniHamed:2007js}
N.~Arkani-Hamed, J.~Orgera, and J.~Polchinski, ``{Euclidean wormholes in string
  theory},'' {\em JHEP} {\bf 0712} (2007) 018,
\href{http://www.arXiv.org/abs/0705.2768}{{\tt 0705.2768}}.

\bibitem{Weinberg:safety}
S.~Weinberg, ``{Critical Phenomena for Field Theorists},''
{\em Lectures presented at Int. School of Subnuclear Physics, Ettore Majorana,
  Erice, Sicily, Jul 23 - Aug 8, 1976}.

\end{thebibliography}\endgroup
\end{document}